\documentclass[%
 reprint,
 amsmath,amssymb,
 aps,
 prd,nofootinbib,nobibnotes,
 superscriptaddress,groupedaddress,
]{revtex4-2}

\usepackage{graphicx}
\usepackage{dcolumn}
\usepackage{bm}
\usepackage[dvipsnames, usenames]{xcolor}
\usepackage{acronym}
\usepackage[hidelinks]{hyperref}
\usepackage{physics}
\usepackage{orcidlink}

\graphicspath{{plots}{plots/posterior_samples}{plots/prior_samples}}

\newcommand{\comment}[1]{}

\newcommand{\ligo}{\affiliation{LIGO Laboratory, Massachusetts Institute of Technology, Cambridge, MA 02139, USA}}
\renewcommand{\mit}{\affiliation{Kavli Institute for Astrophysics and Space Research and Department of Physics, Massachusetts Institute of Technology, Cambridge, MA 02139, USA}}

\begin{document}

\title{PixelPop: High Resolution Nonparameteric Inference of Gravitational-Wave Populations in Multiple Dimensions}

\author{Jack Heinzel\,\orcidlink{0000-0002-5794-821X}}
\email{heinzelj@mit.edu}
\ligo\mit

\author{Matthew Mould\,\orcidlink{0000-0001-5460-2910}}
\ligo\mit

\author{Sofía Álvarez-López\,\orcidlink{0009-0003-8040-4936}}
\ligo\mit

\author{Salvatore Vitale\,\orcidlink{0000-0003-2700-0767}}
\ligo\mit

\date{March 17, 2025}

\begin{abstract}
The origins of merging compact binaries observed by gravitational-wave detectors remains highly uncertain. Several astrophysical channels may contribute to the overall merger rate, with distinct formation processes imprinted on the structure and correlations in the underlying distributions of binary source parameters. In the absence of confident theoretical models, the current understanding of this population mostly relies on simple parametric models that make strong assumptions and are prone to misspecification. Recent work has made progress using more flexible nonparametric models, but detailed measurement of the multidimensional population remains challenging. In pursuit of this, we present \textsc{PixelPop}---a high resolution Bayesian nonparametric model to infer joint distributions and parameter correlations with minimal assumptions. \textsc{PixelPop} densely bins the joint parameter space and directly infers the merger rate in each bin, assuming only that bins are coupled to their nearest neighbors. We demonstrate this method on mock populations with and without bivariate source correlations, employing several statistical metrics for information gain and correlation significance to quantify our nonparametric results. We show that \textsc{PixelPop} correctly recovers the true populations within posterior uncertainties and offers a conservative assessment of population-level features and parameter correlations. Its flexibility and tractability make it a useful data-driven tool to probe gravitational-wave populations in multiple dimensions.
\end{abstract}

\maketitle

\section{Introduction}

The LIGO \cite{LIGOScientific:2014pky}, Virgo \cite{VIRGO:2014yos}, and KAGRA \cite{KAGRA:2020tym} (LVK) gravitational-wave (GW) detectors have observed the mergers of around 100 compact object binaries containing neutron stars and stellar-mass black holes (BHs)~\cite{LIGOScientific:2018mvr,LIGOScientific:2020ibl,LIGOScientific:2021usb,KAGRA:2021vkt}. Though individual detections reveal the nature and properties of these sources such as their masses and spins, GW searches are subject to selection biases, meaning detections are not fair draws from the underlying population of mergers. However, when analyzed collectively, the combined GW catalog can be leveraged to filter out these effects and place constraints on the intrinsic merger rate and astrophysical distribution of sources \cite{Mandel:2018mve, Thrane:2018qnx, Vitale2020, Essick:2023upv}. At the population level, these are ultimately determined by the processes leading to compact binary formation which, at present, face large theoretical uncertainties that GW observations may help constrain.

In particular, different astrophysical formation pathways result in distinct features in the spectra of source parameters \cite{Mandel:2018hfr, Mandel:2021smh, Mapelli:2021taw, Bavera:2020uch, Bavera:2022mef, Zevin:2020gbd, Zevin:2022wrw, Broekgaarden:2022nst, Fuller:2019sxi, Bavera:2020inc, Fuller:2022ysb, Barkat:1967zz, Woosley:2016hmi, Tanikawa:2021zfm, Gerosa:2021mno}. GW observations of binary BH mergers so far imply that \cite{LIGOScientific:2018jsj, LIGOScientific:2020kqk, KAGRA:2021duu}: there is a peak in the merger rate at masses of $\approx10M_\odot$ and a secondary mode between $30M_\odot$ and $40M_\odot$; binaries favor equal mass constituents; at least within the detector horizon, the merger rate increases as a function of redshift; most BH spins are small but not necessarily zero and favor neither orbital alignment nor misalignment strongly. By themselves, these inferences can be compared to theoretical predictions and simulations to determine likely formation pathways that are compatible with observations.

However, population-level measurements are typically made by imposing simplified models for what the true population could be, corresponding to strong prior assumptions. Often these are composed of basic functional forms that are quick and easy to evaluate (power laws, normal distributions, etc.) but do not necessarily have a direct relation to any underlying astrophysics \cite{KAGRA:2021duu}. Another approach is to construct simulation-based models and thereby infer astrophysically relevant parameters directly from GW data \cite{Zevin:2020gbd, Taylor:2018iat, Wong:2020ise, Mould:2022ccw, Riley:2023jep}, though assumptions about the simulated populations can significantly bias analysis of real data \cite{Cheng:2023ddt}. Therefore, on the other end of the spectrum there is growing use of nonparametric methods that sacrifice interpretability for increased model flexibility, including but not limited to: likelihood-maximizing models \cite{Payne:2022xan}; transdimensional models \cite{Toubiana:2023egi}; infinite-dimensional mixtures \cite{Rinaldi:2021bhm}; Gaussian \cite{Mandel:2016prl, KAGRA:2021duu, Ray:2023upk, Farah:2024xub} and autoregressive \cite{Callister:2023tgi} processes; splines \cite{Edelman:2022ydv, Golomb:2022bon}; and more. These relax the stronger prior assumptions of parametric models and allow for more freedom to capture possible features in the population, but also come with their own drawbacks, such as larger measurement uncertainties, computational expense, or difficulty accounting for gravitational wave selection biases.

Moreover, it is often difficult to extend these methods to flexibly model multidimensional distributions. This is crucial because there is important information hiding in the multivariate distribution of source properties rather than the univariate spectra above alone. Astrophysical formation mechanisms may imprint correlations at the population level \cite{PortegiesZwart:2002iks, Rodriguez:2015oxa, Bavera:2020uch, Bavera:2022mef, Santini:2023ukl, Baibhav:2022qxm, Gerosa:2021mno, vanSon:2021zpk, Marchant:2023ncp}---a possibility that the majority of analyses neglect, thus missing key astrophysical insights. Additionally, the presence of multiple subpopulations originating from distinct formation pathways can masquerade as parameter correlations. Indeed, targeted modeling of the joint mass, redshift, and spin distribution of binary BHs observed by the LVK has revealed potential pairwise correlations \cite{Callister:2021fpo, Biscoveanu:2022qac, Franciolini:2022iaa, Li:2023yyt, Pierra:2024fbl, Ray:2024hos, Rinaldi:2023bbd}, though it is important to test that such results are not driven by model choices \cite{Heinzel:2023hlb}. This is precisely the issue we tackle.

We develop a flexible population modeling framework, inspired by analysis of aerial spatial data, to analyse GW populations in multiple dimensions. Our model---\textsc{PixelPop}---makes minimal assumptions, namely only that the merger rate is correlated between neighbouring points in a binned parameter space. We focus on bivariate population distributions and validate our approach on simulated GW catalogs. We demonstrate that \textsc{PixelPop} successfully infers structured correlations in the distributions of binary BH masses, redshifts, and spins, despite assuming little about the nature of the true population.

The remainder of the paper is structured as follows. In Sec.~\ref{sec: Methods} we describe the Bayesian population inference problem and our procedure for estimating the likelihood and handling its uncertainty. In Sec.~\ref{sec: car model} we describe \textsc{PixelPop}. We discuss the importance of using minimal smoothing for astrophysical populations and its computational advantages. In Sec.~\ref{sec: simulated populations} we validate our method on synthetic populations. Finally, we conclude and point to further applications of \textsc{PixelPop} in Sec. \ref{sec:conclusion}. In a companion paper \cite{Heinzel:2024hva}, we analyze in detail possible parameter correlations in the population of binary BH mergers using public LVK data from the third GW transient catalog (GWTC-3). We make available the data for both of these papers at Ref.~\cite{heinzel_2024_13176116}.

\section{Population inference}
\label{sec: Methods}

\subsection{Functional likelihood}
\label{sec: Gravitational wave population likelihood}

Suppose we observe $N_\mathrm{obs}$ GW events with corresponding data $d_n$, $n=1,...,N_\mathrm{obs}$, over an observing period $T_\mathrm{obs}$. We model their occurrence with the detector-frame differential merger rate
\begin{align}
\label{eq: rate density}
R ( \theta )
=
\frac { \dd{N} } { \dd{t_\mathrm{d}} \dd{\theta} }
=
\frac { 1 } { T_\mathrm{obs} }
\frac { \dd{N} } { \dd{\theta} }
\, ,
\end{align}
where $N$ is the total number of mergers occurring over $T_\mathrm{obs}$, $t_\mathrm{d}$ is detector-frame time, and $\theta$ represents (a subset of) the binary source parameters. Equivalently, we may instead consider the source-frame differential merger rate per unit comoving volume $V_\mathrm{c}$ (typically in units $\mathrm{Gpc}^{-3} \, \mathrm{yr}^{-1}$), which is more astrophysically relevant:
\begin{align}
\label{eq: comoving rate density}
\mathcal{R} ( \theta' ; z )
=
\frac
{ \dd{N} }
{ \dd{V_\mathrm{c}} \dd{t_\mathrm{s}} \dd{\theta'} }
=
\left( \frac { 1 } { 1 + z } \dv {V_\mathrm{c}} {z} \right)^{-1}
R ( \theta' , z )
\, ,
\end{align}
where $t_\mathrm{s} = t_\mathrm{d} (1+z)^{-1}$ is the source-frame time, $\theta'$ represents (a subset of) the source parameters excluding redshift $z$, and we use the notation $\mathcal{R} ( \theta' ; z )$ to indicate that $\mathcal{R}$ is not a rate density over redshift.

Modeling GW events as independent draws from an inhomogenous Poisson process results in a Bayesian hierarchical model with the population-level likelihood \cite{Mandel:2018mve, Thrane:2018qnx, Vitale2020, Essick:2023upv}
\begin{align}
\label{eq: likelihood}
\mathcal{L}[R]
=
p ( \{ d_n \} | R )
\propto
e^{ -N_\mathrm{exp} [ R ] }
\prod_{n=1}^{N_\mathrm{obs}} \mathcal{L}_n[R]
\, .
\end{align}
Note that, unlike in other presentations, here $R$ is an infinite dimensional vector (a function) that we model directly, such that $\mathcal{L} [ R ]$ is a functional.
The single-event likelihood functionals are
\begin{align}
\mathcal{L}_n[R]
=
p ( d_n | R )
\propto
\int \dd{\theta_n} p ( d_n | \theta_n ) R ( \theta_n )
\, .
\label{eq: single event likelihood}
\end{align}
The expected number of detections implied by the merger rate model is
\begin{align}
\label{eq: Nexp}
N_\mathrm{exp} [ R ]
=
T_\mathrm{obs}
\iint \dd{d} \dd{\theta}
P ( \mathrm{det} | d )  p ( d | \theta ) R ( \theta )
\, .
\end{align}
Selection biases are accounted for by the probability  $P(\mathrm{det}|d)$ of detecting (``det'') a signal in GW data $d$, typically taken to be an indicator function on the output of search pipelines \cite{KAGRA:2021duu, Essick:2023upv}. The data distribution $p(d|\theta)$, i.e., the likelihood that data $d$ were produced by a source with parameters $\theta$, is determined by the change of variables $d = n + s(\theta)$, where $s$ is the GW signal model and noise $n$ is assumed to be generated by a zero-mean stationary Gaussian process \cite{Finn:1992wt, Cutler:1994ys}.

\subsection{Likelihood estimation}
\label{sec: Likelihood estimation}

Evaluating the GW likelihood $p(d|\theta)$ for each event included in the population-level analysis and marginalizing over their source parameters in Eq.~(\ref{eq: single event likelihood}) is computationally expensive, mostly due to the required waveform evaluations. A common resolution is to instead approximate these integrals with importance sampling, using Bayes' theorem to write $p(d|\theta) \propto p(\theta|d,\mathrm{PE}) / p(\theta|\mathrm{PE})$ for a reference posterior $p(\theta|d,\mathrm{PE})$ inferred under some default parameter-estimation (PE) prior $p(\theta|\mathrm{PE})$. The likelihood for each event is then replaced by an estimator
\begin{align}
\label{eq: single event likelihood estimator}
\hat{\mathcal{L}}_n[R]
=
\frac { 1 } { N_\mathrm{PE} }
\sum_{i=1}^{N_\mathrm{PE}}
\frac { R ( \theta_{ni} ) } { p ( \theta_{ni} | \mathrm{PE} ) }
\, ,
\end{align}
where $N_\mathrm{PE}$ posterior samples are drawn from the initial PE posterior for each event, $\theta_{ni} \sim p(\theta_n|d_n,\mathrm{PE})$.
Similarly, the expected number of detections is estimated as 
\begin{align}
\hat{N}_\mathrm{exp}[R]
=
\frac { T_\mathrm{obs} } { N_\mathrm{inj} }
\sum_{j=1}^{N_\mathrm{det}}
\frac { R ( \theta_j ) } { p ( \theta_j | \mathrm{inj} ) }
\, ,
\label{eq: Nexp estimator}
\end{align}
where $N_\mathrm{inj}$ mock signals that are drawn from a reference distribution $p(\theta|\mathrm{inj})$ result in $N_\mathrm{det}$ signals $d_j,\theta_j \sim P(\mathrm{det}|d) p(d|\theta) p(\theta|\mathrm{inj})$ recovered by the search pipelines \cite{Tiwari:2017ndi,Farr:2019rap}. Altogether, the (natural logarithm of the) population likelihood is replaced with the estimator
\begin{align}
\label{eq: total ln likelihood estimator}
\ln \hat{\mathcal{L}}[R]
=
\sum_{n=1}^{N_\mathrm{obs}}
\ln \hat{\mathcal{L}}_n[R]
-
\hat{N}_\mathrm{exp}[R]
\, .
\end{align}

\subsection{Likelihood uncertainty}
\label{sec: Likelihood uncertainty}

The downside to these estimators is that they carry intrinsic statistical uncertainty due to the finite number of samples used in the Monte Carlo expectation values, which can lead to inferred merger rate distributions with artificially high likelihoods. This is especially problematic for very flexible nonparametric population models, such as the one we propose in Section \ref{sec: car model}. While this can be mitigated by appropriate choices for the reference distributions $p(\theta|\mathrm{PE})$ and $p(\theta|\mathrm{inj})$, the number of drawn samples $N_\mathrm{PE}$ and $N_\mathrm{inj}$, and the merger-rate prior \cite{Essick:2022ojx}, the effect unavoidably grows with increasing catalog size and observing time \cite{Talbot:2023pex}; the total variance of the log-likelihood estimator is
\begin{align}
\hat{\sigma}_{\ln \mathcal{L}}^2 [R]
=
\sum_{n=1}^{N_\mathrm{obs}}
\frac
{ \hat{\sigma}_n^2 }
{ \hat{\mathcal{L}}_n[R]^2 }
+
\hat{\sigma}_\mathrm{exp}^2
\, ,
\label{eq: likelihood variance}
\end{align}
where the variance contributions from the estimators of the single-event likelihoods and expected catalog size are
\begin{align}
\hat{\sigma}_n^2 [R]
& =
\frac { 1 } { N_\mathrm{PE}^2 }
\sum_{i=1}^{N_\mathrm{PE}}
\frac { R(\theta_{ni})^2 } { p(\theta_{ni} | \mathrm{PE} )^2 }
-
\frac { \hat{\mathcal{L}}_n[R]^2 } { N_\mathrm{PE} }
\, ,
\\
\hat{\sigma}_\mathrm{exp}^2 [R]
& =
\frac { T_\mathrm{obs}^2 } { N_\mathrm{inj}^2 }
\sum_{j=1}^{N_\mathrm{det}}
\frac { R(\theta_j)^2 } { p(\theta_j|\mathrm{inj})^2 }
-
\frac { \hat{N}_\mathrm{exp}[R]^2 } { N_\mathrm{inj} }
\, , 
\label{eq: variance terms}
\end{align}
respectively. When the above variances are large, the point estimate in Eq. (\ref{eq: total ln likelihood estimator}) cannot be trusted.

Previous work has prevented high-variance posterior estimates by thresholding the likelihood, sending it to zero when the variances \cite{Talbot:2023pex} or associated effective sample sizes \cite{Farr:2019rap, KAGRA:2021duu} pass ad hoc thresholds. This prevents exploration of such uncertain regions, e.g., by stochastic samplers. Following Ref. \cite{Talbot:2023pex}, we impose $\hat{\sigma}_{\ln\mathcal{L}}^2 < 1$, i.e., that the relative uncertainty in the estimator is less than unity \cite{Essick:2022ojx}. Rather than a strict threshold, we aggressively taper by replacing the likelihood $\ln\mathcal{L}[R] \mapsto \ln\hat{\mathcal{L}}[R] - \mathcal{T}[R]$, where the regularization term is
\begin{align}
\mathcal{T}[R]
=
\begin{cases}
100 ( \hat{\sigma}_{\ln\mathcal{L}}^2 - 1 )^2
& \mathrm{if} \quad \hat{\sigma}_{\ln\mathcal{L}} \geq 1
\, ,
\\
0
& \mathrm{if} \quad \hat{\sigma}_{\ln\mathcal{L}} < 1
\, .
\end{cases}
\label{eq: likelihood taper}
\end{align}
The numerical prefactor is chosen heuristically to be much larger than the intrinsic spread in $\ln\hat{\mathcal{L}}[R]$ (the shape of this function is not a unique choice and other options successfully regularize the likelihood estimator \cite{Callister:2023tgi}). Provided the merger-rate model $R$ is differentiable, the advantage of continuous and differentiable tapering functions $\mathcal{T}$ over step functions is that the likelihood estimator retains differentiability; as we shall see in Section \ref{sec: car model}, this is useful when employing gradient-based sampling.

\section{The PixelPop Model}
\label{sec: car model}

The key ingredient for GW population inference is a model $\mathcal{R}(\theta';z)$ for the source-frame merger rate density. In this work, we aim to model the source distributions as flexibly as possible in multiple dimensions, removing astrophysical and simplified parametric assumptions. To do so, we model the merger rate as piecewise constant by discretizing the parameter space into uniformly spaced bins $\theta_b$, $b=1,...,B$, and inferring the merger rate $\mathcal{R}_b$ in each bin, similar to Refs.~\cite{KAGRA:2021duu, Mandel:2016prl, Vitale:2018yhm, Ng:2020qpk, Ng:2022agi, Ray:2023upk, MaganaHernandez:2024uty, Ray:2024hos}; i.e.,
\begin{align}
\label{eq: binned rate}
\mathcal{R}(\theta)
=
\sum_{b=1}^B
\begin{cases}
\mathcal{R}_b \quad & \mathrm{if} \quad \theta \in \theta_b \, ,
\\
0 \quad & \mathrm{if} \quad \theta \notin \theta_b \, .
\end{cases}
\end{align}
The bins into which each PE sample and found injection fall need to be found just once, so that Eqs.~(\ref{eq: single event likelihood estimator}) and (\ref{eq: Nexp estimator}) can be evaluated.

However, Ref.~\cite{Payne:2022xan} showed that the merger rate model that maximizes the likelihood estimator in Eq.~(\ref{eq: total ln likelihood estimator}) is a weighted sum of delta distributions---a model that is clearly unphysical. To prevent the binned representation converging to an unphysical distribution, we encode our prior belief that the underlying merger rate has some level of continuity over the parameter space with a minimal smoothing prior.

\subsection{Conditional autoregressive prior}
\label{sec: Conditional autoregressive prior}

\begin{figure}
\centering
\includegraphics[width=0.9\linewidth]{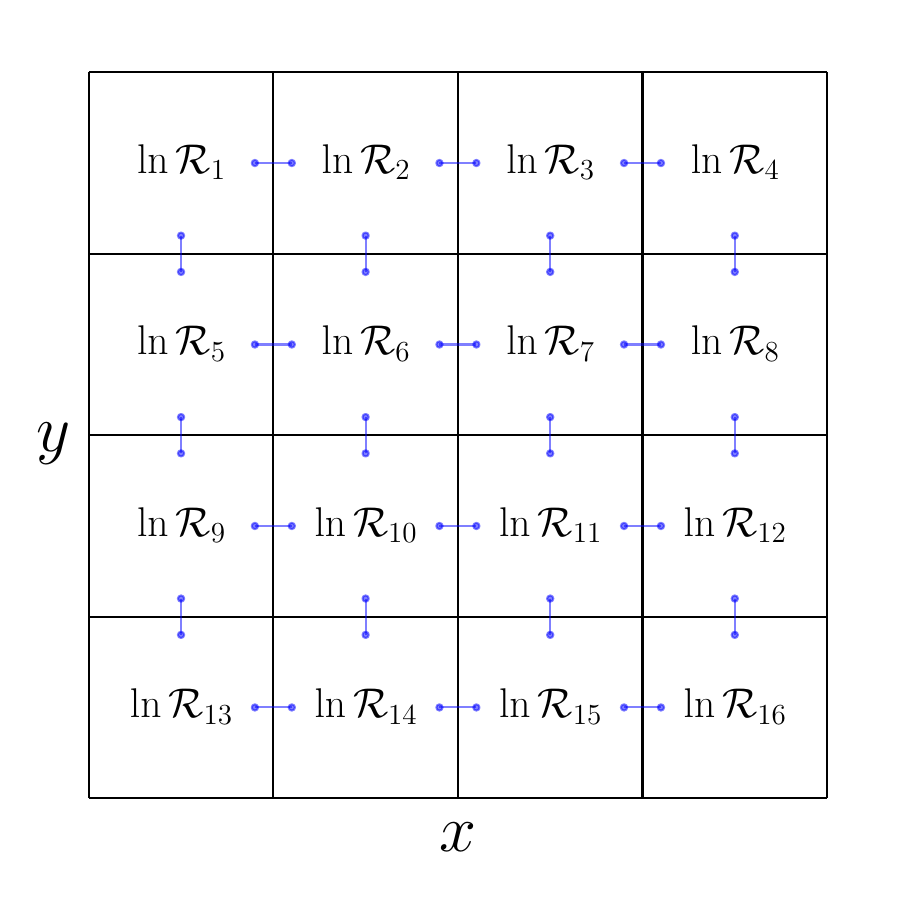}
\caption{A diagram of the bin coupling in two dimensions, with four bins along each axis within which the merger rate density $\mathcal{R}$ is modeled as constant. The blue lines indicate the couplings between bins. The model can be viewed as a graph in which nodes connected by an edge share conditional dependence, while nodes that are not connected are conditionally independent.}
\label{fig: car}
\end{figure}

\textsc{PixelPop} sets a log-normal prior on the comoving merger rate density $\mathcal{R}_b$ (a normal prior in $\ln\mathcal{R}_b$) in each bin. To evaluate the population likelihood, the comoving merger rate density is converted to the detector frame using Eq.~(\ref{eq: comoving rate density}). In this work, we focus on two-dimensional distributions, but in principle binning can occur in any number of dimensions (subject to the curse of dimensionality). Previous studies have used Gaussian process kernels to couple points in parameter space, typically as a function of their continuous or binned separation \cite{Foreman-Mackey_2014, Mandel:2016prl, Vitale:2018yhm, Ng:2020qpk, Ng:2022agi, Ray:2023upk, KAGRA:2021duu, Farah:2024xub}. Instead, we couple bins only to their nearest neighbors, similar in spirit to the one-dimensional autoregressive model of Ref.~\cite{Callister:2023tgi}. This is the weakest possible smoothing prior for the binned multidimensional parameter space while still being sensitive to localized features. The binned coupling is visualized in Fig.~\ref{fig: car}.

\begin{figure*}
\centering

\includegraphics[width=0.4\linewidth]
{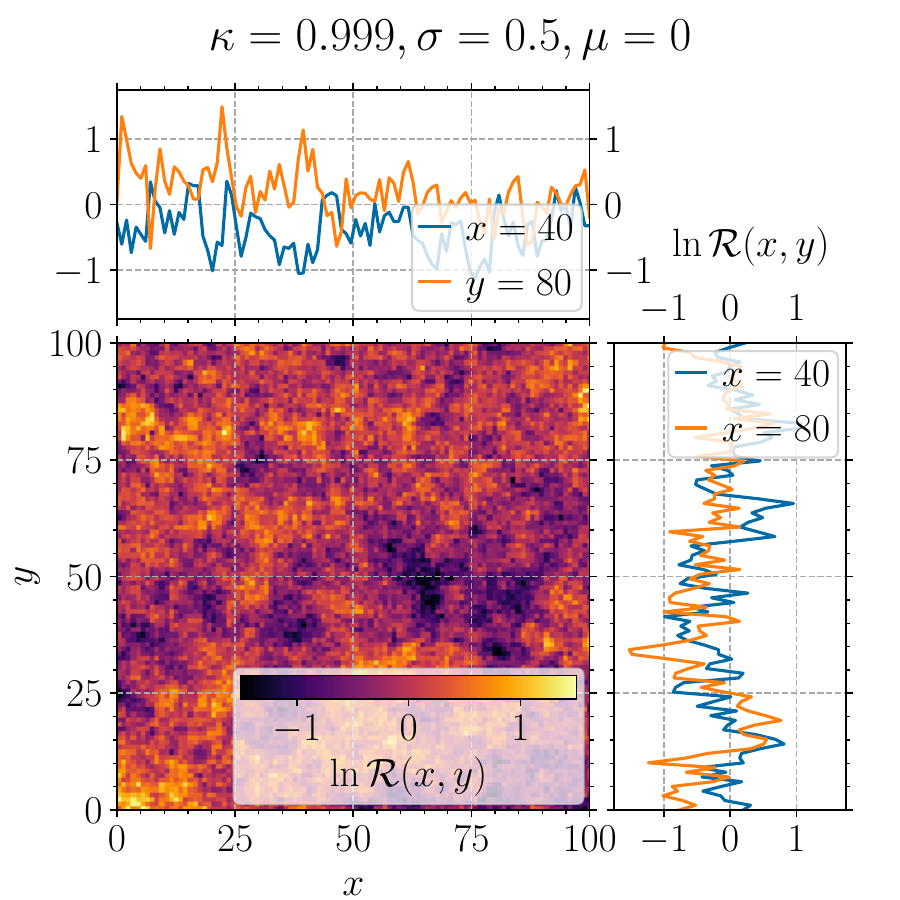}\hspace{6mm}\vspace{-3.5mm}
\includegraphics[width=0.4\linewidth]
{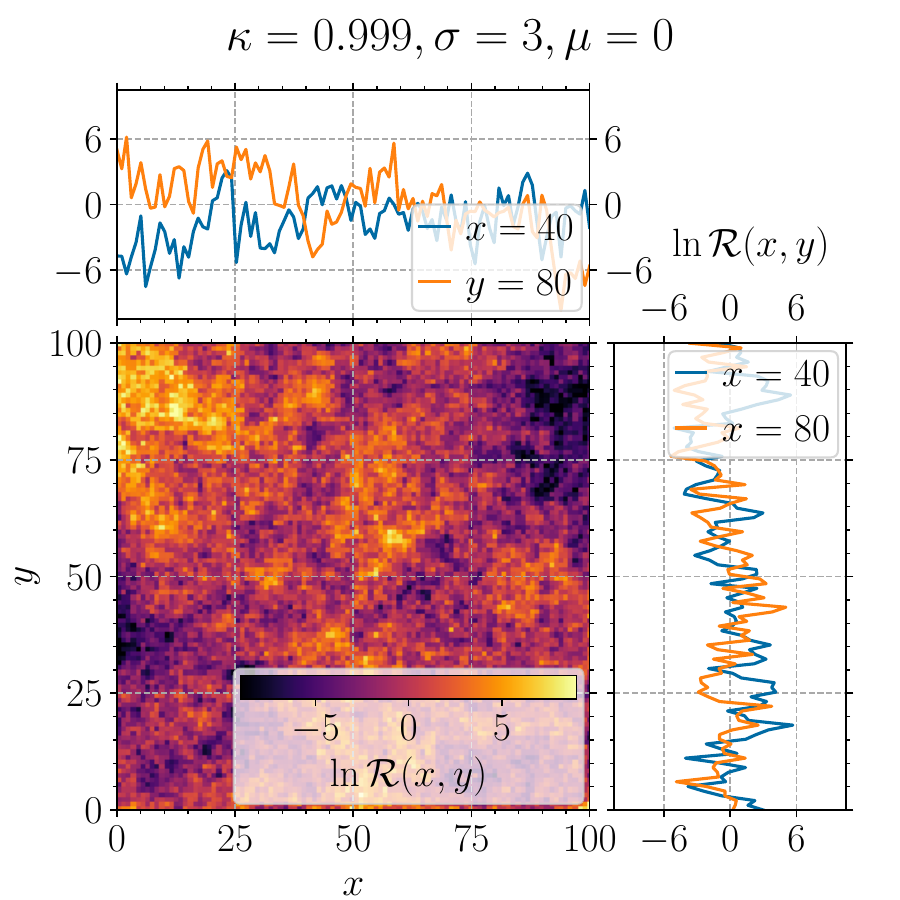}
    
\includegraphics[width=0.4\linewidth]
{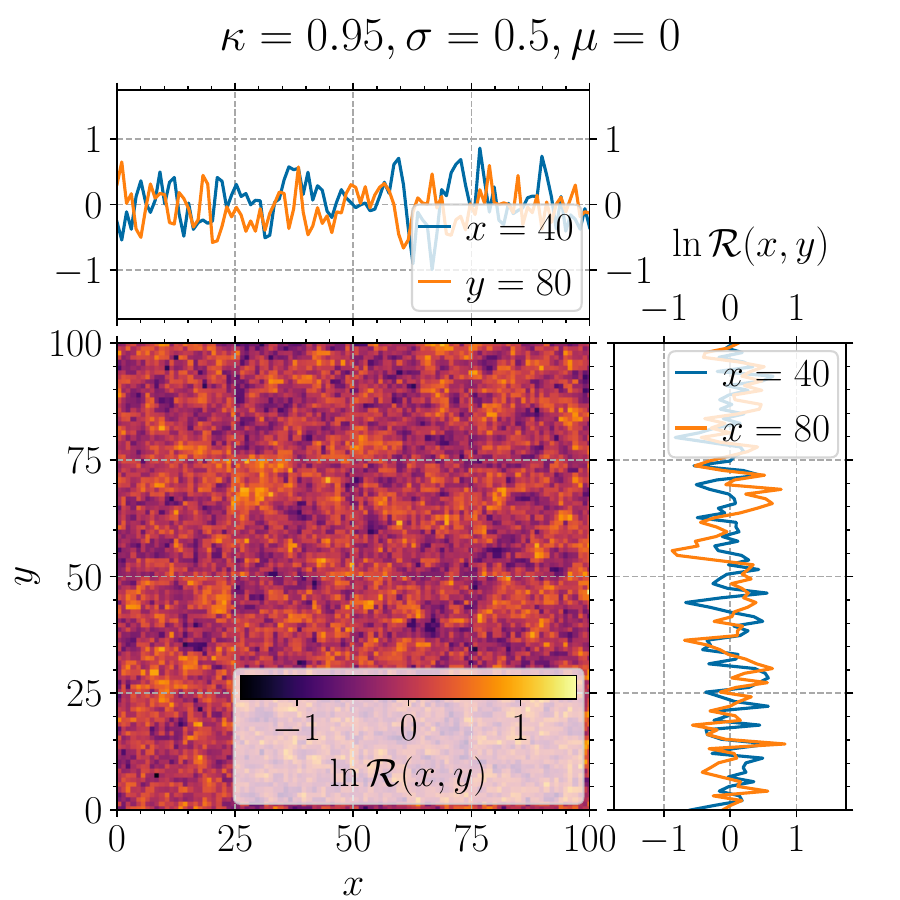} \hspace{6mm}\vspace{-3.5mm}
\includegraphics[width=0.4\linewidth]
{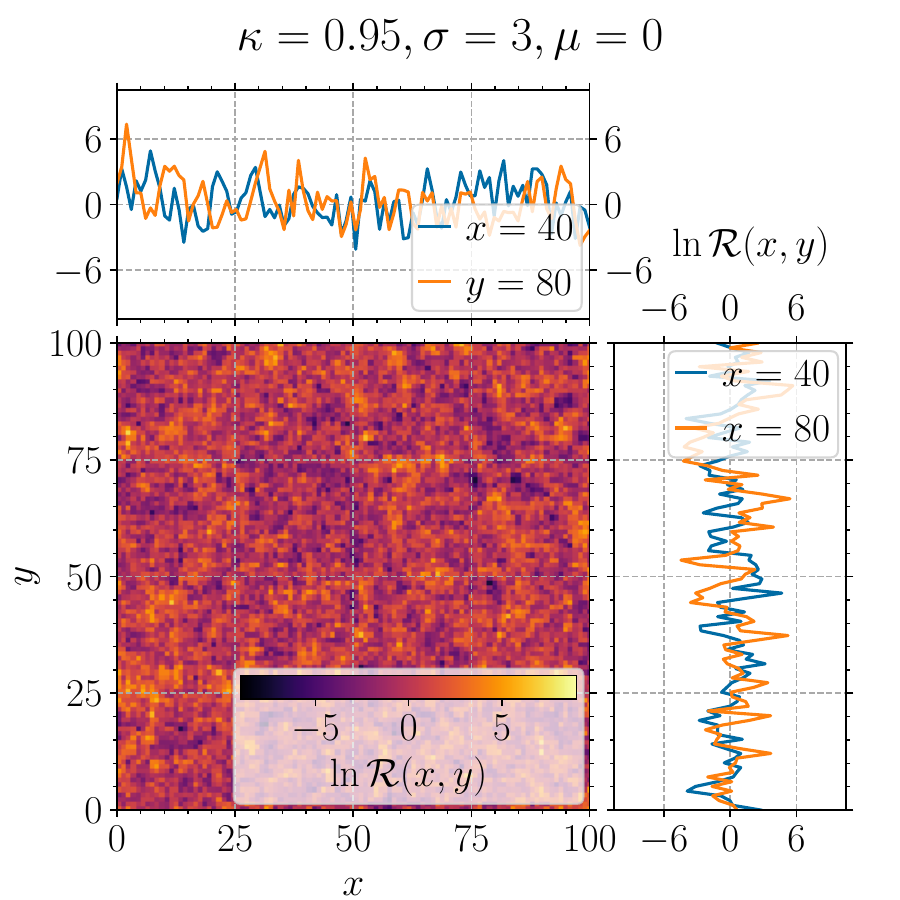}
    
\includegraphics[width=0.4\linewidth]
{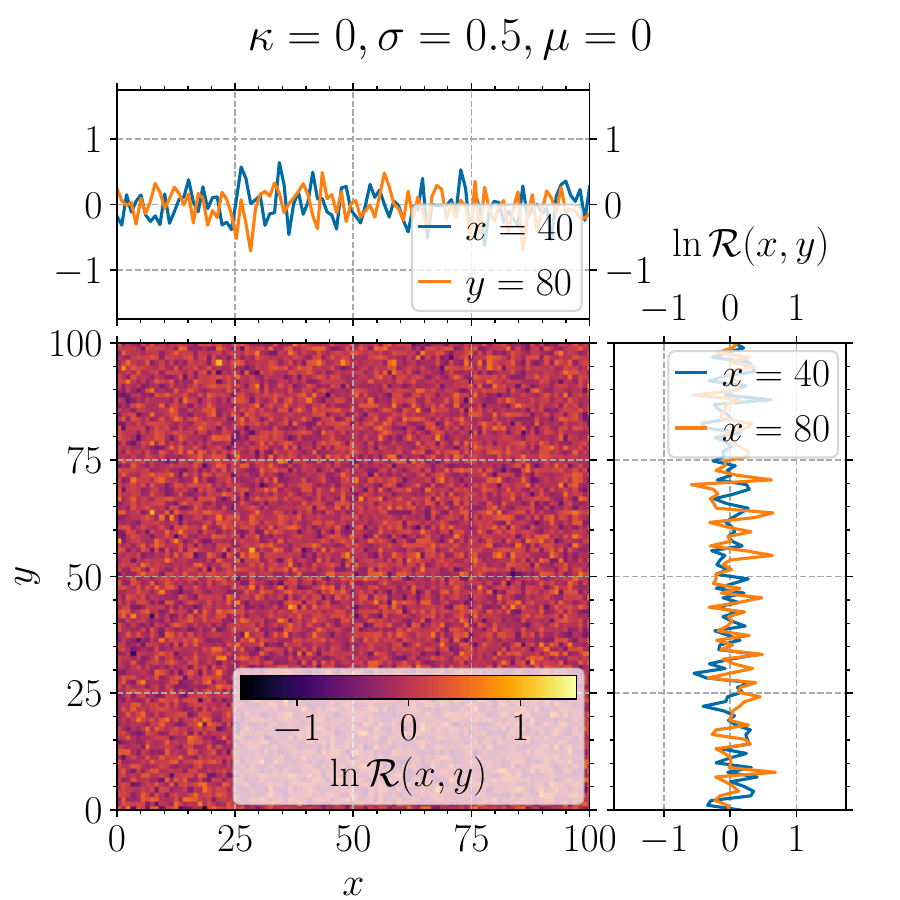}\hspace{6mm}\vspace{-3.5mm}
\includegraphics[width=0.4\linewidth]
{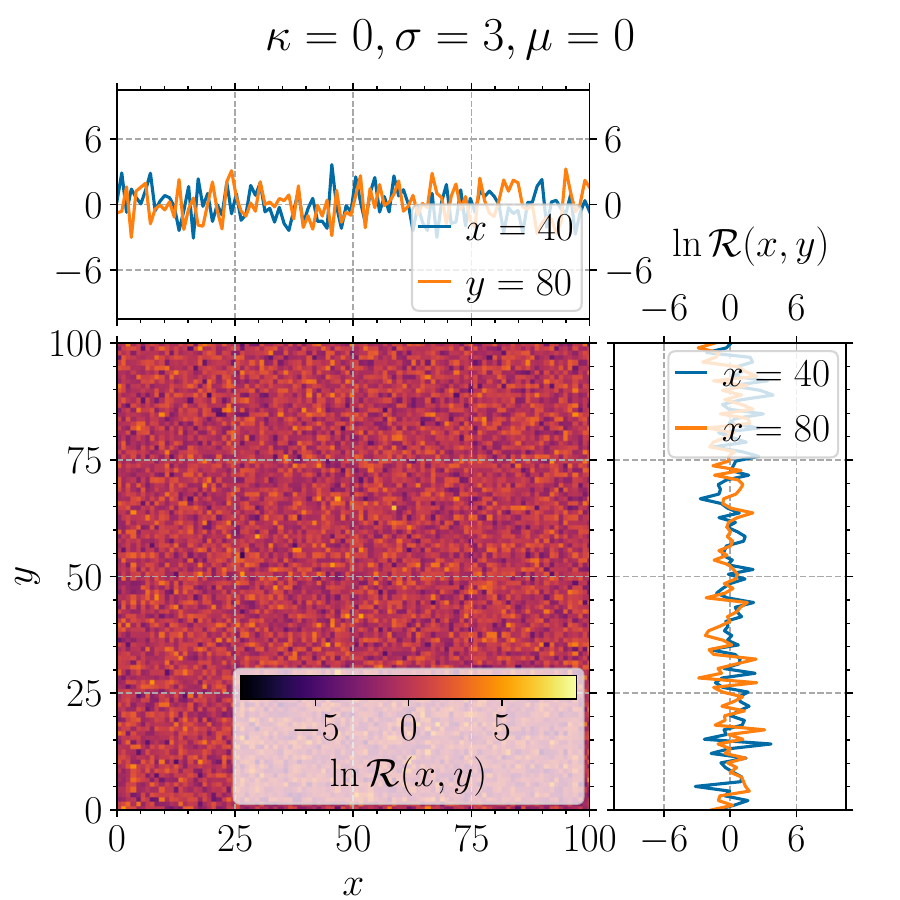}

\caption{
Samples from the CAR model prior for a merger rate density $\ln\mathcal{R}(x,y)$ defined on a two dimensional grid of 100 bins along each axis $x$ and $y$. For each example we show the two-dimensional random sample as a heatmap in the central panel and one-dimensional slices in the upper and right-hand panels. In the left column the global variance parameter $\sigma$ that determines the scale of variability in $\mathcal{R}(x,y)$ is set to $\sigma=0.5$ and in the right column it is set to $\sigma=3$; notice that the range of values of $\ln\mathcal{R}(x,y)$ is larger in the right column. From top to bottom, the local correlation parameter $\kappa$ that sets the strength of coupling between adjacent bins goes from $0.999, 0.95$, to $0$. As $\kappa\to1$ each sample has more localized structure whereas for $\kappa=0$ the prior on each bin becomes an independent normal distribution and the structure is homogeneous. The mean parameter $\mu=0$ is not varied in this figure as it simply shifts the values of $\ln\mathcal{R}(x,y)$ up or down globally.
}
\label{fig: car prior samples}
\end{figure*}

Let $\mathcal{S}_\beta$ denote the set of bins that are immediately adjacent to bin $\theta_\beta$; e.g., in two dimensions, $\mathcal{S}_\beta$ contains four bins that share an edge with bin $\theta_\beta$ if it is not a boundary or corner bin, three bins if $\theta_\beta$ lies on a single boundary of the parameter space, and two if it is a corner bin; see Fig.~\ref{fig: car}. The conditional prior on $\ln\mathcal{R}_\beta$ is
\begin{align}
\label{eq: car example}
& p ( \ln\mathcal{R}_\beta | \{ \ln\mathcal{R}_{b\neq\beta} \} , \kappa , \sigma , \mu )
\\
& =
p ( \ln\mathcal{R}_\beta | \{ \ln\mathcal{R}_{b\in\mathcal{S}_b} \} , \kappa , \sigma , \mu )
\nonumber \\
& \propto
\exp[
- \sum_{b \in \mathcal{S}_\beta}
\frac
{ \kappa ( \ln\mathcal{R}_\beta - \ln\mathcal{R}_b )^2 + ( 1 - \kappa ) ( \ln\mathcal{R}_\beta - \mu )^2 }
{ 2 \sigma^2 }
]
\nonumber
\, .
\end{align}
The global scale of $\mathcal{R}$ is set by the mean $\mu$ and the global variation in the allowed values of $\mathcal{R}$ is set by $\sigma$. The coupling between bin $\beta$ and its nearest neighbors is weighted by the correlation parameter $0\leq\kappa\leq1$, while coupling to the global mean $\mu$ is weighted by $1-\kappa$  (in principle one can allow $-1\leq\kappa\leq1$, with negative values leading to a ``repulsion'' between the rates in neighboring bins). When $\kappa=0$ the conditional prior in Eq.~(\ref{eq: car example}) reduces to independent normal distributions for each bin, while for $\kappa=1$ there is conditional dependence between adjacent bins only. This is known as a conditional autoregressive (CAR) model and is often used in the analysis of spatial data \cite{Cressie2015-ii, Banerjee2011-ym, Rue2023-kt, DeOliveira2010}. Writing the set of all rates as a vector $\boldsymbol{\mathcal{R}} = [\mathcal{R}_1 \ ... \ \mathcal{R}_B]^\mathrm{T}$, their joint CAR prior distribution is given by \cite{Besag:1974abc, Besag:1995car}
\begin{align}
& p ( \{ \ln\mathcal{R}_b \} | \kappa , \sigma , \mu )
=
\sqrt{
\frac
{ \det ( \mathbf{D} - \kappa \mathbf{A} ) }
{ (2\pi \sigma^2 )^B }
}
\nonumber \\
& \times
\mathrm{exp}
\bigg[
- ( \ln\boldsymbol{\mathcal{R}} - \mu \mathbf{I} )^\mathrm{T}
\frac { \mathbf{D} - \kappa \mathbf{A} } { 2 \sigma^2 }
( \ln\boldsymbol{\mathcal{R}} - \mu \mathbf{I} )
\bigg]
\, .
\label{eq: car prior}
\end{align}
Here, $\mathbf{I}$ is the $B$-dimensional identity vector, $\mathbf{A}$ is the adjacency matrix---a sparse symmetric $B \times B$ matrix to mask bins that are not nearest neighbors---and $\mathbf{D}$ is a diagonal matrix that counts the number of nearest neighbors for each bin \cite{WHITE20093033}, i.e.,
\begin{align}
& A_{ii} = 0
\, , \quad
A_{ij} = A_{ji}
=
\begin{cases}
1 \quad \mathrm{if} \ \theta_i , \theta_j \ \mathrm{adjacent} \, ,
\\
0 \quad \mathrm{otherwise} \, ,
\end{cases}
\label{eq: adjacency matrix}
\\
& D_{ii} = \sum_{j=1}^B A_{ij}
\, , \quad
D_{ij} = 0 \ (i \neq j)
\, .
\label{eq: neighbor matrix}
\end{align}
In Fig.~\ref{fig: car prior samples} we show some example draws from the CAR model for different values of the coupling parameters $\kappa$ and $\sigma$ that control the structure of the generated distributions; the mean $\mu$ serves only as a global shift. While $\kappa$ determines how much local structure there is, $\sigma$ sets the overall range of values that $\ln\mathcal{R}$ can take.

For a generic multivariate normal distribution, evaluating properly normalized densities requires computing both the inverse and determinant of the covariance matrix. When setting a normal prior over $B$ merger-rate bins, the $B \times B$ covariance matrix results in an unfavorable $\mathcal{O}(B^3)$ computational complexity. This is a severe issue for Gaussian process methods that have been used for GW population inference \cite{KAGRA:2021duu, Mandel:2016prl, Vitale:2018yhm, Ng:2020qpk, Ng:2022agi, Ray:2023upk, MaganaHernandez:2024uty, Ray:2024hos, Farah:2024xub, Li:2021ukd}. Especially in higher dimensions, this forces a compromise between the number of bins and the computational feasibility---reducing the bin count allows the model to be evaluated in a reasonable time, but at the cost of losing resolution across the parameter space. 

The form of the CAR prior in Eq.~(\ref{eq: car prior}) has several computational advantages over other covariance kernels for Gaussian process priors. The precision matrix $(\mathbf{D} - \kappa \mathbf{A}) / \sigma^2$ of this normal distribution---that is, the inverse of the covariance matrix---is modeled directly, such that the covariance matrix never needs to be inverted. Its determinant can be rapidly computed for generic values of $\kappa$ and $\sigma$ using the decomposition
\begin{align}
\det [ ( \mathbf{D} - \kappa \mathbf{A} ) / \sigma^2 ]
=
\det ( \mathbf{D} ) \det ( \mathbf{I} - \kappa \mathbf{D}^{-1} \mathbf{A} ) / \sigma^{2B}
\, .
\end{align}
Since $\mathbf{D}$ is a diagonal matrix the first determinant is trivial: $\det(\mathbf{D}) = \prod_{i=1}^B D_{ii}$. The second determinant is proportional to the characteristic polynomial of the scaled adjacency matrix $\mathbf{D}^{-1}\mathbf{A}$, meaning we need only compute its eigenvalues once. This reduces the computational scaling of the determinant calculation to $\mathcal{O}(B)$.

In summary, \textsc{PixelPop} offers the dual advantages of greatly reduced computational complexity while simultaneously imposing weaker assumptions about the form of the merger rate density, as generated by the CAR model. We can therefore probe the multidimensional parameter space with much higher resolution, a crucial advantage that makes \textsc{PixelPop} more sensitive to GW populations with complicated structures and parameter correlations.

\subsection{Posterior and sampling}
\label{sec: Posterior and sampling}

\textsc{PixelPop} simultaneously samples the merger rate densities $\mathcal{R}_b$ in each bin, $b=1,...,B$, as well as the parameters $\Omega=(\kappa,\sigma,\mu)$ of the CAR model. Using the likelihood in Eq.~(\ref{eq: likelihood}) and Bayes' theorem, the joint hierarchical posterior distribution is
\begin{align}
\label{eq: posterior}
p ( \{\mathcal{R}_b\} , \Omega | \{d_n\} )
\propto
p ( \{d_n\} | \{\mathcal{R}_b\} )
p ( \{\mathcal{R}_b\} | \Omega )
p ( \Omega )
\, ,
\end{align}
where the likelihood $p ( \{d_n\} | \{\mathcal{R}_b\} )$ is evaluated using Eqs.~(\ref{eq: total ln likelihood estimator}) and (\ref{eq: likelihood taper}), for which the merger rate density is evaluated using Eq.~(\ref{eq: binned rate}). We set independent priors for the CAR parameters $p(\Omega)=p(\kappa)p(\sigma)p(\mu)$. We take a uniform prior on $\ln(1-\kappa)$ for $\kappa\in[0,1)$, a uniform prior $\ln\sigma\in[-3,5]$, and a uniform prior on $\mu\in[-50,100]$. In general, we find that $\kappa$ favors values $\approx1$. In this limit, the merger-rate bins do not couple to the global mean $\mu$ and as such $\mu$ does not affect the likelihood and sampling it just returns the prior; see Eq.~(\ref{eq: car example}). The variance $\sigma$ in the values of the merger-rate density is constrained away from the prior, however.

For the two-dimensional merger-rate density models we consider, we take a default of 100 bins per dimension. Since for GW population analyses we are typically interested in the population distribution of at least five source parameters---two masses, two (effective) spins, and redshift---we complement \textsc{PixelPop} with standard parametric models for the remaining dimensions it does not fit. This means that the posterior in Eq.~(\ref{eq: posterior}) has at least $10^4$ parameters. To efficiently sample such a high-dimensional distribution we use Hamiltonian Monte Carlo (HMC) sampling \cite{Duane:1987de, Neal:2011mrf, Betancourt:2017ebh}, which has much better computational scaling with the dimensionality of the target posterior compared to other Markov Chain Monte Carlo (MCMC) methods. HMC is a gradient-based sampling method that requires derivatives of the posterior density, meaning the likelihood and prior model must both be differentiable (as alluded to in Sec.~\ref{sec: Likelihood uncertainty}). We use the \texttt{NumPyro} \cite{Phan:2019elc} implementation of the No-U-Turn Sampler (NUTS) \cite{Hoffman:2011ukg} which leverages automatic differentiation within the \texttt{JAX} framework \cite{jax2018github} to compute derivatives of the target posterior for HMC.

\subsection{Quantifying correlations}
\label{sec: spearman}

As \textsc{PixelPop} is a nonparametric model that directly infers the merger rate density $\mathcal{R}$ through its binned representation $\{\mathcal{R}_b\}_{b=1}^B$, we cannot immediately interpret features or correlations in the inferred population, unlike for parametric models. As an example, if we were to assume that BH masses are normally distributed, the mean and variance would be model parameters that could be measured with a population analysis and directly inform us of the location and scale of the mass distribution. If we were to allow the mean mass to linearly depend on redshift, we could infer the slope of the relationship and use its posterior to directly assess whether there is or is not a positive or negative population-level correlation between mass and redshift \cite{Callister:2021fpo, Biscoveanu:2022qac, Safarzadeh:2020mlb}. Instead, with \textsc{PixelPop} we must use the dependence of the merger rate density on the source parameters through the inferred posterior in each bin. While we can visually inspect the inferred merger rate density bins to identify features, we require a quantitative metric to assess their significance.

For two random variables $x$ and $y$, a measure of linear correlation between them is the Pearson correlation coefficient,
\begin{align}
\rho_\mathrm{p} ( x , y )
=
\frac
{ \mathrm{Cov}(x,y) }
{ \sqrt{ \mathrm{Var}(x) \mathrm{Var}(y) } }
\in
[ -1 , 1 ]
\, ,
\label{eq: Pearson}
\end{align}
where $\mathrm{Cov}(x,y)$ is the covariance between $x$ and $y$, and $\mathrm{Var}(x)$, $\mathrm{Var}(y)$ are their variances. If $x$ and $y$ are exactly linearly correlated with a positive (negative) slope, then $\rho_\mathrm{p}(x,y) = 1$ ($-1$) regardless of the magnitude of the slope, while $\rho_\mathrm{p}(x,y)=0$ implies $x$ and $y$ have no linear dependency. We would like to quantify possible nonlinear correlations, however. To do so, we can instead consider the order and rank statistics of $x$ and $y$; the $k$th order statistic of a statistical sample is its $k$th smallest value, while the ranks are the indices of the order statistics. For example, for a sample $x=\{15,9,14\}$, the ranks are $\mathrm{Rank}(x)=\{3,1,2\}$ and the 1st, 2nd, and 3rd order statistics are $\{9,14,15\}$. The Spearman rank correlation coefficient \cite{Spearman:1904abc, kendall1979advanced, Schweizer1981} is defined as
\begin{align}
\rho_\mathrm{s} ( x , y )
=
\rho_\mathrm{p} \big( \mathrm{Rank}(x) , \mathrm{Rank}(y) \big)
\in
[ -1 , 1 ]
\, ,
\label{eq: Spearman}
\end{align}
which is symmetric. A monotonic but nonlinear relationship between $x$ and $y$ implies a linear relationship between $\mathrm{Rank}(x)$ and $\mathrm{Rank}(y)$, such that we can use the Pearson correlation coefficient on the rank variables to quantify nonlinear correlation. If there is a perfect monotonic---but not necessarily linear---relationship, with $x$ increasing (decreasing) as $y$ increases, then $\rho_\mathrm{s}(x,y)=1$ ($-1$). If there is no correlation between $x$ and $y$, then $\rho_\mathrm{s}(x,y)=0$. See Ref.~\cite{Hauke:2011abc} for a comparison between the Pearson and Spearman correlation coefficients.

The Spearman correlation coefficient quantifies monotonic correlations between source parameters, but not other kinds of correlations. We may, for instance, wish to quantify the significance that the distribution of one parameter broadens as a function of another while remaining symmetric. For example, for a constant-mean normal distribution of BH masses for which the width is an increasing but nonlinear function of redshift, the Spearman rank correlation coefficient between mass and redshift would be unsuitable because $\rho_\mathrm{s}=0$. Instead, we use a ``broadening coefficient'' between two source parameters $x$ and $y$,
\begin{align}
\rho_\mathrm{b}(x,y)
=
\rho_\mathrm{s} \left( x , [ y - \mathrm{E}(y) ]^2 \right)
\, ,
\label{eq: broadening}
\end{align}
which is the Spearman rank correlation coefficient between $x$ and the squared deviations $[y-\mathrm{E}(y)]^2$ of $y$, where $\mathrm{E}(y)$ denotes the expectation value of $y$, and is not a symmetric function of $x$ and $y$. Note that the squared deviations are themselves random variables, unlike, say, the variance. If $\rho_\mathrm{b}(x,y)$ is positive (negative), the distribution of $y$ broadens (narrows) as a function of $x$, while it is zero if there is no correlation.

Since \textsc{PixelPop} provides a binned representation of the merger rate density, we can readily compute a properly normalized probability density by renormalization: if each bin is uniformly spaced and has volume $\delta\theta$, the probability density is
\begin{align}
p(\theta)
=
\frac
{ \mathcal{R}(\theta) }
{ \int \dd{\theta'} \mathcal{R}(\theta') }
=
\frac
{ \mathcal{R}(\theta) }
{ \delta\theta \sum_{b=1}^B \mathcal{R}(\theta_b) }
\propto
\mathcal{R}(\theta)
\, .
\label{eq: binned probability density}
\end{align}
The correlation statistic $\rho_\mathrm{s}$ can then be computed with a Monte Carlo approximation:
\begin{enumerate}
\item randomly sample a bin $\theta_b$ according to the density in Eq.~(\ref{eq: binned probability density});
\item randomly sample parameter values $\theta$ uniformly within the bin $\theta_b$;
\item repeat the first two steps to draw a large statistical sample of parameter values $\{\theta\} \sim p(\theta)$;
\item numerically order from lowest to highest each of the one-dimensional samples $\{x\}$ and $\{y\}$ for the subset of parameters $x,y\in\theta$ of interest and compute their ranks, $\{\mathrm{Rank}(x)\}$ and $\{\mathrm{Rank}(y)\}$;
\item compute the sample variances of $\{\mathrm{Rank}(x)\}$ and $\{\mathrm{Rank}(y)\}$ and the sample covariance between them; and
\item use Eqs.~(\ref{eq: Pearson}) and (\ref{eq: Spearman}) to compute the correlation coefficient $\rho_\mathrm{s}(x,y)$.
\end{enumerate}
To compute the broadening statistic, at step 4 above compute the sample mean $\hat{y}$ and continue with the squared deviations $\{(y-\hat{y})^2\}$ instead of the samples $\{y\}$, using Eqs.~(\ref{eq: Pearson}--\ref{eq: broadening}) to compute $\rho_\mathrm{b}(x,y)$.

From the posterior of the merger rate densities in Eq.~(\ref{eq: posterior}), we can then infer posterior distributions for the correlation and broadening coefficients. As we shall see in Sec.~\ref{sec: simulated populations}, we argue that it is the signs of the correlation and broadening coefficients---not their magnitudes---that offer robust evidence for or against correlations. There is a very strong prior against $\rho_\mathrm{s},\rho_\mathrm{b}=\pm1$ simply because there are many more ways to make uncorrelated or slightly correlated distributions than there are perfectly correlated distributions. Note that the choice of using squared deviations in the definition of the broadening statistic is unique up to any nonlinear but strictly increasing transformation because the Spearman correlation coefficient depends only on the corresponding rank variables; e.g., using the absolute deviations instead would result in the same value of $\rho_\mathrm{b}$.

\subsection{Quantifying information gain}
\label{sec: Quantifying information gain} 

We would also like to quantify how informative the GW data are across the space of source parameters. To marginalize out the influence of the CAR model parameters $\Omega=(\kappa,\sigma,\mu)$, we define the marginal posterior and an effective ``informed prior'' for the binned merger rate densities as
\begin{align}
\mathcal{P} ( \{\mathcal{R}_b\} )
& =
\int \dd{\Omega}
p ( \{\mathcal{R}_b\} , \Omega | \{d_n\} )
\, ,
\label{eq: posterior all bins}
\\
\pi ( \{\mathcal{R}_b\} )
& =
\int \dd{\Omega}
p ( \{\mathcal{R}_b\} | \Omega ) p ( \Omega | \{d_n\} )
\label{eq: effective prior all bins}
\, ,
\end{align}
respectively. Whereas the former marginalizes the full posterior from Eq.~(\ref{eq: posterior}) over the CAR parameters $\Omega$, the latter marginalizes the conditional prior $p(\{\mathcal{R}_b\}|\Omega)$ over the posterior of the CAR parameters only. Similarly, we can compute the one-dimensional posteriors and effective priors for a single bin $b=\beta$ as
\begin{align}
\mathcal{P} ( R_\beta )
& =
\int \dd { \{ \mathcal{R}_{b\neq\beta} \} }
\mathcal{P} ( \{ \mathcal{R}_b \} )
\, ,
\label{eq: posterior single bin}
\\
\pi ( \mathcal{R}_\beta )
& =
\int \dd { \{ \mathcal{R}_{b\neq\beta} \} }
\pi ( \{ \mathcal{R}_b \} )
\nonumber \\
& =
\iint \dd{\Omega} \dd { \{ \mathcal{R}_{b\neq\beta} \} }
p ( \{\mathcal{R}_b\} | \Omega ) p ( \Omega | \{d_n\} )
\, ,
\label{eq: effective prior single bin}
\end{align}
respectively. The Kullback--Leibler (KL) divergence between posterior and effective prior,
\begin{align}
\mathrm{KL} [ \mathcal{P}(\mathcal{R}_\beta) , \pi(\mathcal{R}_\beta) ]
=
\int \dd{\mathcal{R}_\beta}
\mathcal{P}(\mathcal{R}_\beta)
\log_2 \frac
{ \mathcal{P}(\mathcal{R}_\beta) }
{ \pi(\mathcal{R}_\beta) }
\, ,
\label{eq: kl}
\end{align}
gives the information gained in each merger rate density bin, where we use base-2 logarithm such that the KL divergence has units of bits. We estimate one-dimensional posterior densities $\mathcal{P}(\mathcal{R}_\beta)$ using a Gaussian kernel density estimate (KDE) fit to posterior samples for $\mathcal{R}_\beta$, drawn as described in Sec.~\ref{sec: Posterior and sampling}. For the effective prior, the inner integral over the other bins $\{\mathcal{R}_{b\neq\beta}\}$ in Eq.~(\ref{eq: effective prior single bin}) can be computed in closed form following Eq.~(\ref{eq: car prior}), while the outer integral is computed as the sample mean over posterior samples for the CAR parameters $\Omega$, which again are drawn as described in Sec.~\ref{sec: Posterior and sampling}.

\section{Simulated Populations}
\label{sec: simulated populations}

\begin{table*}
\renewcommand{\arraystretch}{1.25}
\centering
\begin{tabular}{lllll}
\hline\hline
Parameter & Description & $q$--$\chi_\mathrm{eff}$ correlation & $z$--$\chi_\mathrm{eff}$ correlation & No correlation \\
\hline $\alpha$ & $m_1$ power-law index & $3$ & $3$ & $3$ \\
$\beta_q$ & $q$ power-law index & 1 & 1 & 1 \\
$m_\mathrm{min}$ & minimum BH mass & $5M_\odot$ & $5M_\odot$ & $5 M_\odot$ \\
$m_\mathrm{max}$ & maximum BH mass & $85M_\odot$ & $85M_\odot$ & $85 M_\odot$ \\
$\lambda_\mathrm{peak}$ & fraction of BBHs in Gaussian component & $0.03$ & $0.03$ & $0.03$ \\
$\mu_m$ & $m_1$ Gaussian component mean & $35M_\odot$ & $35 M_\odot$ & $35 M_\odot$ \\
$\sigma_m$ & $m_1$ Gaussian component standard deviation  & $5M_\odot$ & $5M_\odot$ & $5M_\odot$ \\
$\delta_m$ & low-mass smoothing parameter & $3M_\odot$ & $3M_\odot$ & $3M_\odot$ \\
\hline
$\kappa$ & $z$ power-law index & 2 & 2 & 2 \\ \hline
$(x_0, \mu_{\chi_\mathrm{eff}:0})$ & first mean spline node coordinates & $(0, 0.4)$ & $(0, 0)$ & $\mu_{\chi_\mathrm{eff}} = 0.06$ \\
$(x_1, \mu_{\chi_\mathrm{eff}:1})$ & second mean spline node coordinates & $(0.4, 0.3)$ & $(0.3, 0)$ & \\
$(x_2, \mu_{\chi_\mathrm{eff}:2})$ & third mean spline node coordinates & $(0.8, 0.05)$ &  $(0.65, 0)$ & \\
$(x_3, \mu_{\chi_\mathrm{eff}:3})$ & fourth mean spline node coordinates & $(1, 0.02)$ & $(2.3, 0)$ & \\
$(x_0, \ln\sigma_{\chi_\mathrm{eff}:0})$ & first standard deviation spline node coordinates & $(0, -2.5)$& $(0, -3.5)$  & $\ln\sigma_{\chi_\mathrm{eff}} = -2.2$ \\
$(x_1, \ln\sigma_{\chi_\mathrm{eff}:1})$ & second standard deviation spline node coordinates & $(0.4, -2.5)$ & $(0.3, -2)$ & \\
$(x_2, \ln\sigma_{\chi_\mathrm{eff}:2})$ & third standard deviation spline node coordinates & $(0.8, -2.5)$ & $(0.65, -1.5)$ &\\
$(x_3, \ln\sigma_{\chi_\mathrm{eff}:3})$ & fourth standard deviation spline node coordinates & $(1, -2.5)$ & $(2.3, -1.25)$ &\\
\hline\hline
\end{tabular}
\caption{Parameters for the simulated populations, with descriptions and numerical values. The first section is for the \textsc{Power Law + Peak} mass model, the second for the \textsc{Power Law} redshift model, and the third is for the Gaussian model in effective spin $\chi_\mathrm{eff}$---whose mean $\mu_{\chi_\mathrm{eff}}$ and standard deviation $\sigma_{\chi_\mathrm{eff}}$ can be correlated with other source parameters with a cubic spline function. The spline node placements are given by the $(x_i, \mu_{\chi_\mathrm{eff}:i})$ and $(x_i, \mu_{\ln\sigma_\mathrm{eff}:i})$ coordinate pairs, where $x$ is either binary mass ratio $q$ or redshift $z$. For the population in which $q$ and $\chi_\mathrm{eff}$ are correlated (third right-most column), $\mu_{\chi_\mathrm{eff}}$ increases as $q$ decreases, while $\sigma_{\chi_\mathrm{eff}}$ is constant. For the population in which the $z$ is correlated with $\chi_\mathrm{eff}$ (second right-most column), $\sigma_{\chi_\mathrm{eff}}$ increases as $z$ increases, while $\mu_{\chi_\mathrm{eff}}$ is constant. For the population with no parameter correlations (right-most column), both $\mu_{\chi_\mathrm{eff}}$ and $\sigma_{\chi_\mathrm{eff}}$ are constant. The fourth population we consider is an equal mixture of detections from the $q-\chi_\mathrm{eff}$ and $z-\chi_\mathrm{eff}$ correlated populations.
}
\label{tab:true_hyper}
\end{table*}

We test the robustness of \textsc{PixelPop} with four custom populations, modeling the populations of the heavier (primary) BH mass $m_1$, the binary mass ratio $q\in(0,1]$, redshift $z$, and effective aligned spin \cite{Racine:2008qv}
\begin{align}
\chi_\mathrm{eff}
=
\frac
{ \chi_1 \cos\theta_1 + q \chi_2 \cos\theta_2 }
{ 1 + q }
\in (-1,1)
\, ,
\end{align}
where $\chi_1,\chi_2\in[0,1)$ are the dimensionless spin magnitudes of the primary and secondary BH components and $\theta_1,\theta_2$ are their spin--orbit misalignment angles. Two of the four populations we consider have a bivariate correlation, one with a correlation between $q$ and $\chi_\mathrm{eff}$ and another in which the distribution of $\chi_\mathrm{eff}$ broadens over $z$. Third, we consider a control population in which no correlations are present. Finally, we investigate the impact of unmodelled correlations by considering a fourth population with a \textit{trivariate} correlation, while we only model bivariate correlations with \textsc{PixelPop}. The correlations are described in more detail in the following subsections. In each case we take the true mass and redshift distributions to be the \textsc{Power Law + Peak} mass model \cite{Talbot:2018cva} and \textsc{Power Law} redshift model \cite{Fishbach:2018edt}; their parameters are given in Tab.~\ref{tab:true_hyper} and also see App.~B of Ref.~\cite{KAGRA:2021duu}.

For all four populations we consider catalogs of 400 GW events detected over a period of two years by a two-detector LIGO network (LIGO Hanford and LIGO Livingston), assuming power spectral densities (PSDs) representative of the fourth LVK observing run (O4) \cite{KAGRA:2013rdx}. We draw sources from the true populations, generate GW signals using the \textsc{IMRPhenomXP} waveform approximant \cite{Pratten:2020ceb}, and add the detector-projected signals to Gaussian noise colored by each PSD. We consider a source as detected and add it to the catalog if it has a network matched-filter signal-to-noise ratio (SNR) $>9$ \cite{Essick:2023toz, Mould:2023eca}.

We note that this approximation is not consistent with the detection model and likelihood described in Sec.~\ref{sec: Gravitational wave population likelihood} because it assumes detection depends on both the GW data $d$ and the true source parameters $\theta$, rather than $d$ alone as in real GW searches \cite{Essick:2023upv}, i.e., $P(\mathrm{det}|d,\theta) \neq P(\mathrm{det}|d)$. To be self consistent, the population likelihood in Eq.~(\ref{eq: single event likelihood}) should be modified as
\begin{align}
p(d|R)
\propto
\int \dd{\theta}
P(\mathrm{det|d,\theta}) p(d|\theta) R(\theta)
\, ,
\end{align}
which, for our choice of detection criterion, masks the likelihood $p(d|\theta)$ for values of the source parameters $\theta$ that result in a network matched-filter SNR $<9$. This would reduce the effective number of PE samples in the Monte Carlo estimator for the single-event likelihoods of Eq.~(\ref{eq: single event likelihood estimator}) and thus increase the overall uncertainty in the likelihood estimator, as discussed in Sec.~\ref{sec: Likelihood uncertainty}. For this reason, we choose to neglect this modification. However, even with 400 events in our simulated catalogs, we find that any systematic bias introduced is obfuscated by statistical uncertainty.

For each event in our mock catalogs, we perform full PE to draw posterior samples of the source parameters used for Eq.~(\ref{eq: single event likelihood estimator}). We use the same waveform approximant above and the heterodyning/relativing-binning scheme of Refs.~\cite{Cornish:2010kf, Cornish:2021lje, Zackay:2018qdy} to speed up evaluations of the likelihood $p(d|\theta)$, as implemented in \textsc{Bilby} \cite{Ashton:2018jfp, Krishna:2023bug}. We set uniform priors over detector-frame component masses and spin magnitudes, isotropic priors for spin directions and sky location, a redshift prior that is uniform in comoving volume, and uninformative uniform priors for the remaining parameters (e.g., isotropic sky location). The component spin priors are converted to the induced prior on effective spin $\chi_\mathrm{eff}$ following Ref.~\cite{Callister:2021gxf}.

We create an additional set of software injections for computing Eq.~(\ref{eq: Nexp estimator}). For the reference distribution $p(\theta|\mathrm{inj})$ we use the same mass and redshift distributions as the true population (see Tab.~\ref{tab:true_hyper}). We draw spins uniform in magnitude and isotropic in direction. All remaining source parameters follow the uninformative PE priors. We draw a total of $5\times10^8$ sources, $10^6$ of which are detectable according to our SNR criterion. Note that we do not perform PE for these sources as we only require computing the SNR to determine detectability.

\subsection{Mass ratio and effective spin correlation}
\label{sec: Mass ratio and effective spin correlation}

\citeauthor{Callister:2021fpo}~\cite{Callister:2021fpo} showed that there is evidence for an anticorrelation between effective spin $\chi_\mathrm{eff}$ and mass ratio $q$ in the population of real LVK binary BH mergers, and this was confirmed in Refs.~\cite{KAGRA:2021duu, Adamcewicz:2022hce, Adamcewicz:2023mov, Heinzel:2023hlb}. The inferred form of the correlation is subject to strong assumptions for the population, however. We test whether \textsc{PixelPop} can accurately infer such parameter correlations---which may have nontrivial forms---with the mock population from \citeauthor{Heinzel:2023hlb}~\cite{Heinzel:2023hlb}, in which the distributions of $\chi_\mathrm{eff}$ and $q$ have a nonlinear correlation. In particular, the population of $\chi_\mathrm{eff}$ follows a normal distribution truncated on $[-1,1]$ with constant standard deviation $\sigma_{\chi_\mathrm{eff}}$ but a mean $\mu_{\chi_\mathrm{eff}}$ that depends on $q$ with a cubic spline. The cubic spline has four nodes placed at $q = 1, 0.8, 0.4, 0$, over which the $\chi_\mathrm{eff}$ mean increases through $\mu_{\chi_\mathrm{eff}} = 0.02, 0.05, 0.3, 0.4$; see Tab.~\ref{tab:true_hyper}. While we do not expect such a population would be exactly realised in nature, it is representative of complicated astrophysical source distributions for which strongly parametrized models would fail and for which \textsc{PixelPop} is well suited.

We use \textsc{PixelPop} to infer the two-dimensional merger rate density $\mathcal{R}(q,\chi_\mathrm{eff})$, making no assumptions about the functional form of the joint distribution. The dependence on all the source parameters we model is given by
\begin{align}
\mathcal{R} ( m_1 , q , \chi_\mathrm{eff} ; z )
=
\mathcal{R} ( q, \chi_\mathrm{eff} )
p ( m_1 | \Lambda ) p( z | \Lambda)
\, ,
\label{eq: model q chieff}
\end{align}
where we simultaneously infer the parameters $\Lambda$ of the primary mass $m_1$ and redshift $z$ distributions using the parametric \text{Power Law + Peak} mass model (this models both $m_1$ and $q$, so we use just the $m_1$ part of the model here) and \textsc{Power Law} redshift model, rather than fixing them to the true distributions. The parameters of these models are described in Tab.~\ref{tab:true_hyper}. Note that the redshift-dependent prefactor in Eq.~(\ref{eq: comoving rate density}) is included in the redshift model; see App.~{B\,3} of Ref.~\cite{KAGRA:2021duu}. We draw $10^4$ posterior samples, following Sec.~\ref{sec: Posterior and sampling}.

\begin{figure}
\centering
\includegraphics[width=1\linewidth]{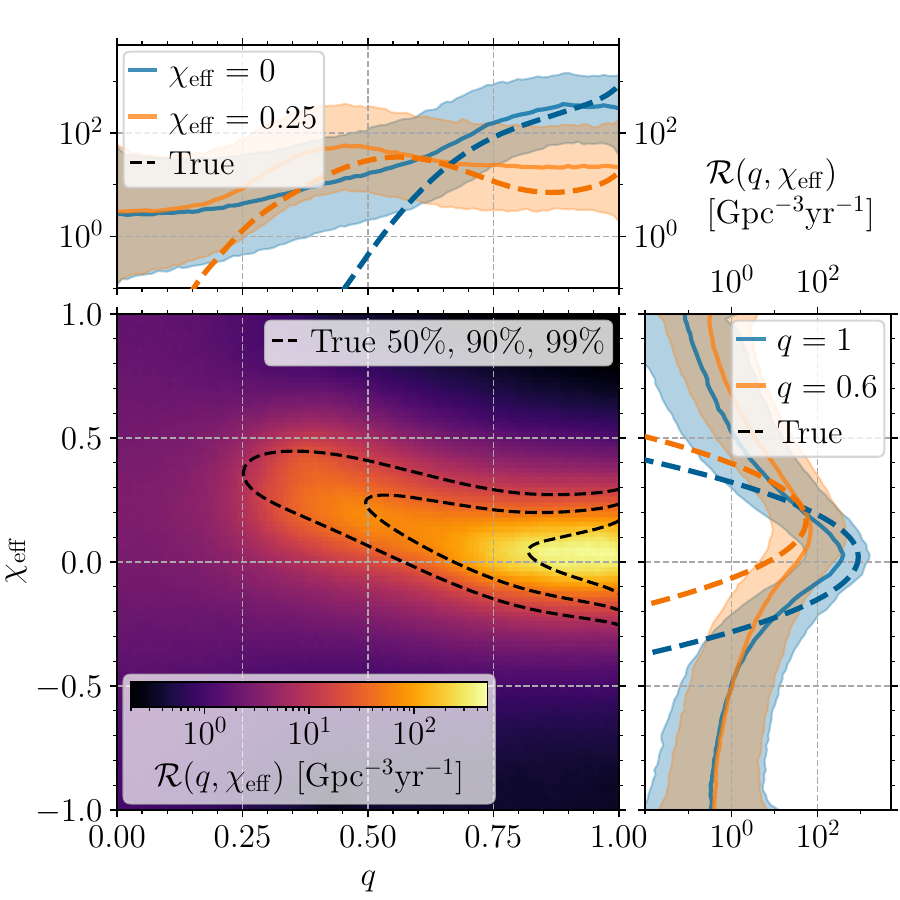}
\caption{Inferred comoving merger rate density $\mathcal{R}(q,\chi_\mathrm{eff};z=0.2)$, evaluated at a fixed redshift $z=0.2$, for the simulated population in which there is a correlation between binary mass ratio $q$ and effective spin $\chi_\mathrm{eff}$. The central panel displays the two-dimensional posterior median, with lower to higher values shaded darker to brighter. The underlying true population is visualized with dashed black lines enclosing 50\%, 90\%, and 99\% of the distribution. The upper panel shows slices $\mathcal{R}(q,\chi_\mathrm{eff}=0;z=0.2)$ (blue) and $\mathcal{R}(q,\chi_\mathrm{eff}=0.25;z=0.2)$ (orange) of the merger rate density as a function of $q$ at fixed values of $\chi_\mathrm{eff}$. The solid lines gives the posterior median and the shaded regions encloses the 90\% posterior credible regions. The dashed lines give the marginal distributions of the true population. Similarly, the right-hand panel shows slices $\mathcal{R}(q=1,\chi_\mathrm{eff};z=0.2)$ (blue) and $\mathcal{R}(q=0.6,\chi_\mathrm{eff};z=0.2)$ (orange) as a function of $\chi_\mathrm{eff}$ for fixed values of $q$.}
\label{fig:qchi}
\end{figure}

In Fig. \ref{fig:qchi}, we show the inferred comoving merger rate density $\mathcal{R}(q,\chi_\mathrm{eff};z=0.2)$, evaluated at a redshift of $z=0.2$ and marginalized over the mass distribution. We include the parametric redshift model evaluated at a particular value of the redshift because $\mathcal{R}$ is the comoving merger rate density, which implicitly depends on redshift through the comoving volume element. The central panel shows the median value of $\mathcal{R}$ in each bin, compared to the 50\%, 90\%, and 99\% credible regions of the true $q$--$\chi_\mathrm{eff}$ population. It is clear that larger merger rate densities, indicated by the brighter shading, trace the true population, meaning \textsc{PixelPop} successfully infers the underlying correlation---despite making no assumptions about it or that there even is one.

The median represents a marginalization of the distribution though, and not the full posterior uncertainty. Due to the difficulty of visualizing uncertainty in two-dimensional distributions, we show that posterior uncertainties for the one-dimensional marginal merger rate densities in the upper and right-hand panels of Fig.~\ref{fig:qchi}. The solid lines and shaded regions show the medians and 90\% credible regions, respectively, while the dashed lines show the marginal distributions of the true underlying population. In regions where the merger rate density is high, \textsc{PixelPop} correctly recovers the true population, within the posterior uncertainty. However, for small rates our result deviates from the true underlying distribution. This is a generic difference between nonparametric models like \textsc{PixelPop} and parametric models: the latter enforce strong assumptions like the population density must become very small toward the tails of parameter space, while for \textsc{PixelPop} there is not enough information in the GW data to distinguish between merger rates of, e.g., $\mathcal{R}(q,\chi_\mathrm{eff}) = 10^{-2} \mathrm{Gpc}^{-3} \, \mathrm{yr}^{-1}$ and $10^{-10} \mathrm{Gpc}^{-3} \, \mathrm{yr}^{-1}$, both of which would be consistent with the lack of detections in those regions of parameter space. The CAR prior favors broader distributions in the absence of informative data, an affect seen in Fig.~\ref{fig:qchi} for $q\ll1$ and $|\chi_\mathrm{eff}|\gg0$.

\begin{figure}
\centering
\includegraphics[width=0.9\columnwidth]{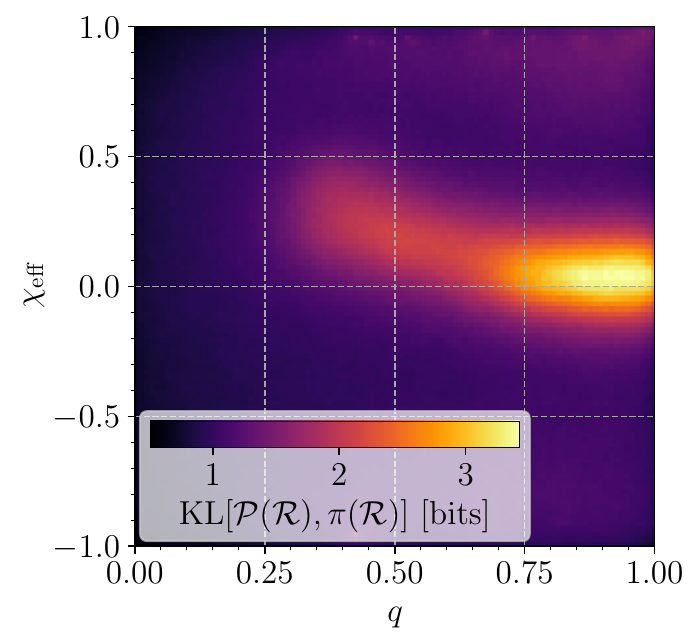}
\caption{Information gained by the posterior $\mathcal{P}(\mathcal{R})$ inferred by \textsc{PixelPop} for each merger rate density pixel over the effective prior $\pi(\mathcal{R})$, defined in Sec.~\ref{sec: Quantifying information gain}, for the population with a nonlinear correlation between binary mass ratio $q$ and effective spin $\chi_\mathrm{eff}$. The information gain is quantified with the KL divergence $\mathrm{KL}[\mathcal{P}(\mathcal{R}),\pi(\mathcal{R})]$ in units of bits, with higher (lower) values shaded brighter (darker).}
\label{fig: kl q chieff}
\end{figure}

Next, we compute the information gained by the posterior for each merger rate density bin $\mathcal{P}(\mathcal{R})$ over the effective prior $\pi(\mathcal{R})$, as defined in Sec.~\ref{sec: Quantifying information gain}. Using the KL divergence $\mathrm{KL}[\mathcal{P}(\mathcal{R}),\pi(\mathcal{R})]$ from Eq.~(\ref{eq: kl}), we plot the result in Fig.~\ref{fig: kl q chieff}. As above, the merger rate densities are evaluated at $z=0.2$. For bins which are brighter (darker), the inferred posterior differs more from (is more similar to) the prior. By comparing against Fig.~\ref{fig:qchi}, we see that larger values of the KL divergence trace the area of larger merger rate densities inferred by \textsc{PixelPop}. On the other hand, in the regions of lower inferred merger rate around $q\approx1$ and $|\chi_\mathrm{eff}|\approx1$, the KL divergence does not decrease as much because sources with $q\approx1$ are on average more \textit{detectable} than others, but for $|\chi_\mathrm{eff}|\approx1$ have not been \textit{detected}, implying the underlying merger rate must be lower. The combination of nondetection and knowledge of GW selection effects informs the posterior.

\begin{figure}
\centering
\includegraphics[width=\linewidth]{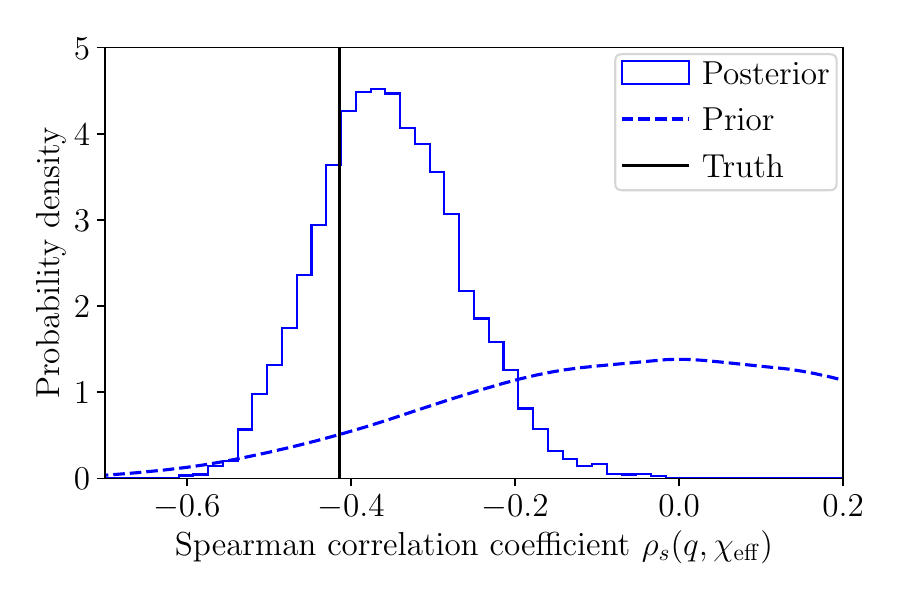}
\caption{Spearman correlation coefficient $\rho_\mathrm{s}(q,\chi_\mathrm{eff})$ from Eq.~(\ref{eq: Spearman})) between the binary mass ratio $q$ and effective spin $\chi_\mathrm{eff}$, measured for $q\in[0.2,1]$, for the synthetic population in which there is a nonlinear correlation between $q$ and $\chi_\mathrm{eff}$. The solid line shows the posterior, while the dashed line shows the effective prior induced by the CAR model when marginalized over posterior of the CAR parameters; see Eq.~(\ref{eq: effective prior all bins}). The value $-0.41$ for the true population is given by the vertical black line.}
\label{fig:spearman_qchi}
\end{figure}

Finally, we quantify the evidence for a correlated $q$--$\chi_\mathrm{eff}$ population using the Spearman correlation coefficient $\rho_\mathrm{s}(q,\chi_\mathrm{eff})$ defined in Eq.~(\ref{eq: Spearman}). We restrict mass ratios to the range $q\in[0.2,1]$ to avoid prior dominated regions of parameter space. The true value for the synthetic population is $\rho_\mathrm{s}(q,\chi_\mathrm{eff})=-0.41$. We show the posterior for $\rho_\mathrm{s}(q,\chi_\mathrm{eff})$ in Fig.~\ref{fig:spearman_qchi} and compare it to the prior induced by the effective prior from Eq.~(\ref{eq: effective prior all bins}), which visibly disfavors values $|\rho_\mathrm{s}(q,\chi_\mathrm{eff})| \approx 1$. The posterior is clearly constrained away from the prior and is fully consistent with the true value. We find $\rho_\mathrm{s}(q,\chi_\mathrm{eff}) = -0.36^{+0.15}_{-0.14}$ (median and 90\% credibility); moreover, $\rho_\mathrm{s}(q,\chi_\mathrm{eff}) < 0$ for all posterior samples, saturating the precision possible with $10^4$ samples---a very confident identification of a negative correlation between $q$ and $\chi_\mathrm{eff}$.

In summary, our analysis suggests that after O4, when the real GW catalog may be of a size similar to the mock catalog here \cite{KAGRA:2013rdx}, we can use \textsc{PixelPop} to \textit{confidently} identify $q$--$\chi_\mathrm{eff}$ correlations \textit{without any strong parametric assumptions}.

\subsection{Redshift and effective spin correlation}
\label{sec: Redshift and effective spin correlation}

\begin{figure*}
\centering
\includegraphics[width=0.42\linewidth]{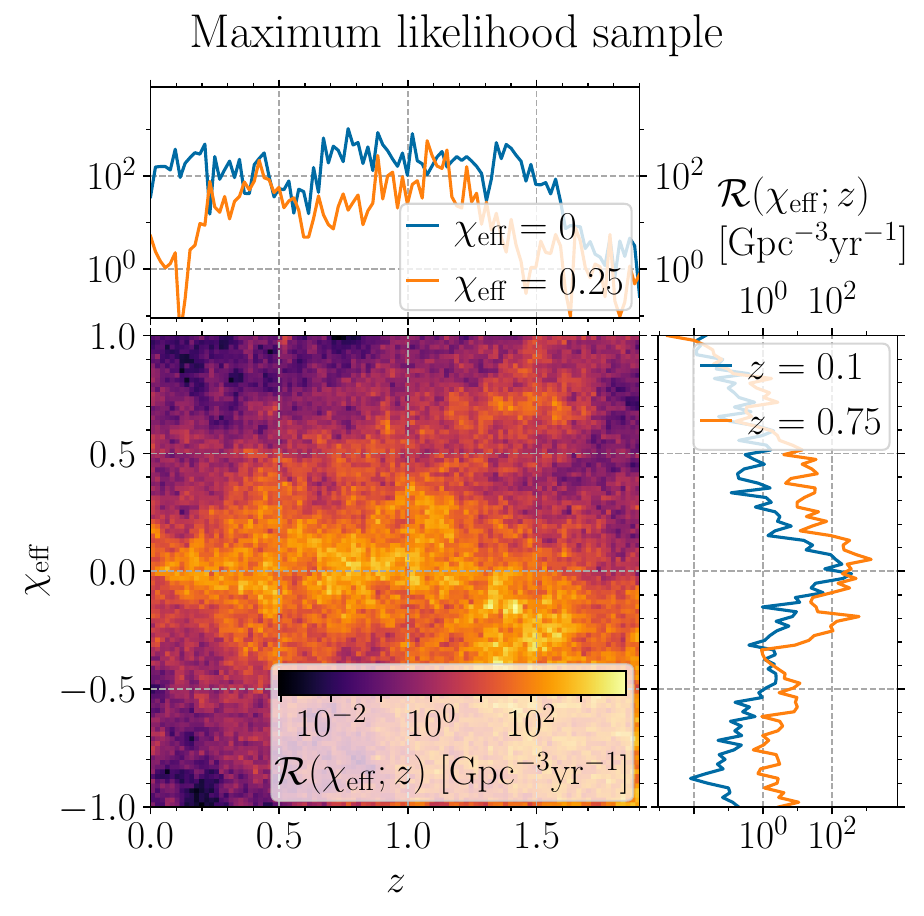}
\hspace{6mm}
\includegraphics[width=0.42\linewidth]{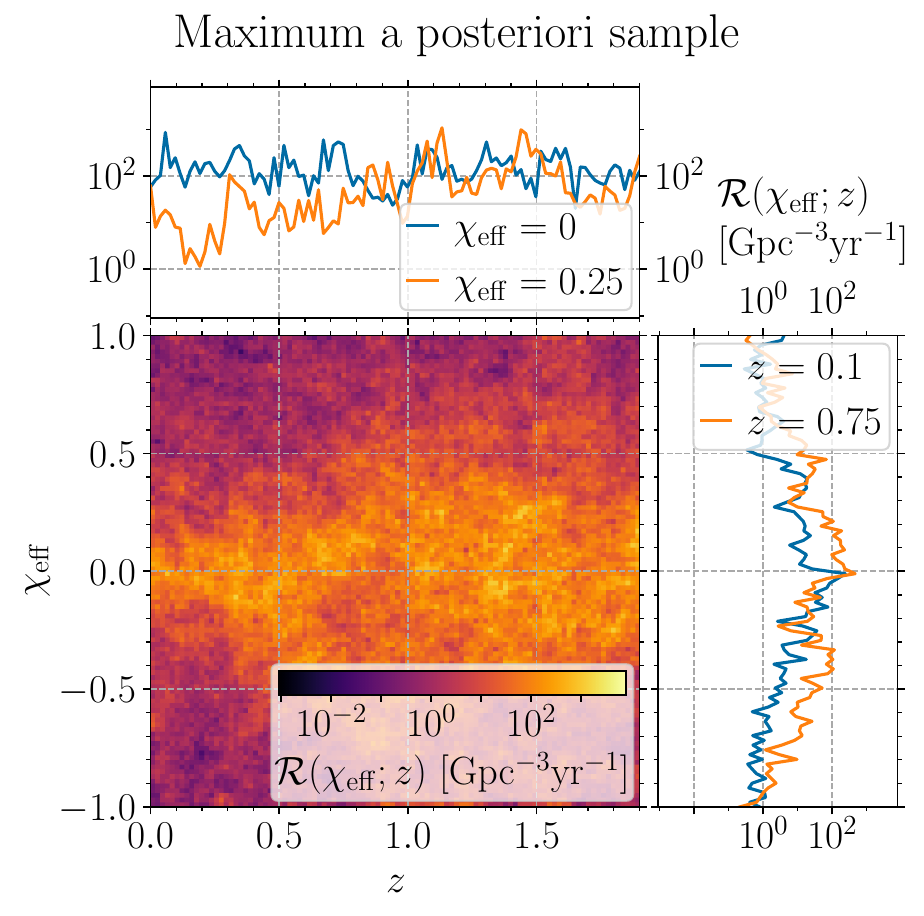}
\caption{
Two posterior samples of \textsc{PixelPop} from inference on the population in which the effective spin $\chi_\mathrm{eff}$ is correlated with redshift $z$. The left and right plots show the maximum likelihood and maximum a posteriori samples, respectively. In each case, the central panel shows the full two-dimensional merger rate density bin posteriors, while the upper panels show evaluations at fixed $\chi_\mathrm{eff}=0$ (blue) and $\chi_\mathrm{eff}=0.25$ (orange) and the right-hand panels at fixed $z=0.1$ (blue) and $z=0.75$ (orange).}
\label{fig:zchi_posterior_samples}
\end{figure*}

\begin{figure}
\centering
\includegraphics[width=1\linewidth]{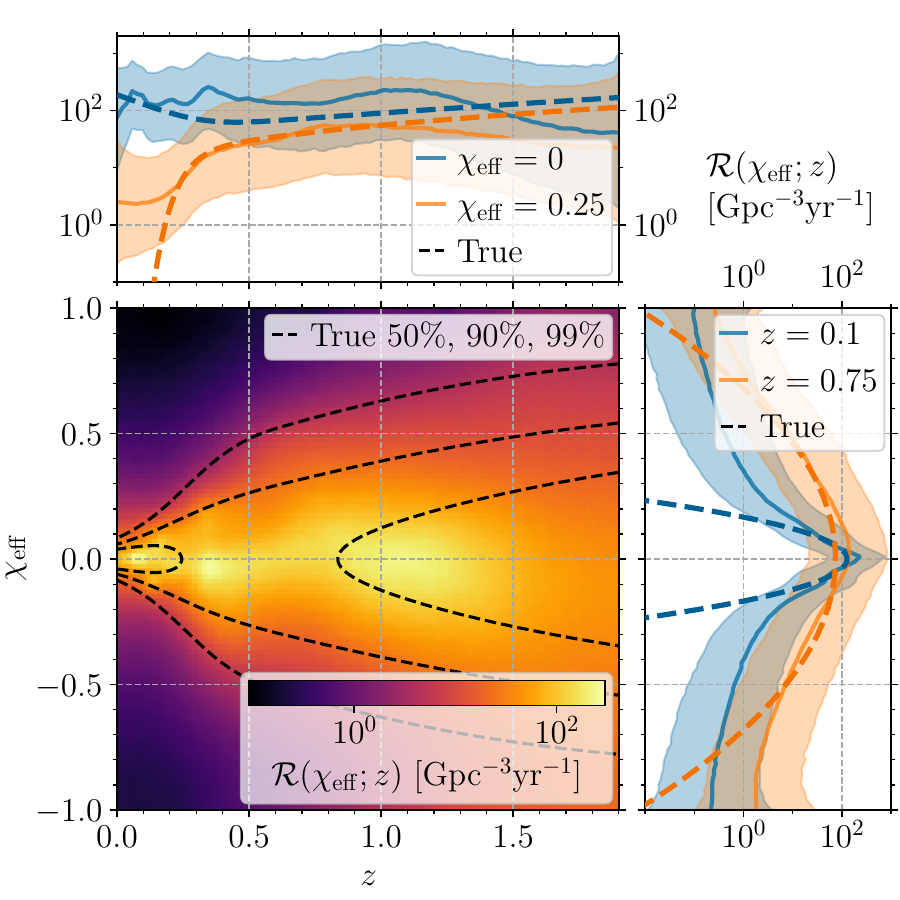}
\caption{Inferred comoving merger rate density $\mathcal{R}(\chi_\mathrm{eff};z)$ for a simulated population in which there is a correlation between redshift $z$ and effective spin $\chi_\mathrm{eff}$. The central panel displays the two-dimensional posterior median and dashed black line enclose 50\%, 90\%, and 99\% of the true distribution. The upper panel shows slices $\mathcal{R}(\chi_\mathrm{eff}=0;z)$ (blue) and $\mathcal{R}(\chi_\mathrm{eff}=0.25;z)$ (orange). The solid lines give the posterior medians and the shaded regions enclose the 90\% posterior credible regions. The dashed lines give the marginal distributions of the true population. Similarly, the right-hand panel shows slices $\mathcal{R}(\chi_\mathrm{eff};z=0.1)$ (blue) and $\mathcal{R}(\chi_\mathrm{eff};z=0.75)$ (orange).}
\label{fig:zchi}
\end{figure}

\citeauthor{Biscoveanu:2022qac}~\cite{Biscoveanu:2022qac} found evidence for a correlation between redshift $z$ and effective spin $\chi_\mathrm{eff}$ in the LVK binary BH population. In particular, the width of the $\chi_\mathrm{eff}$ distribution likely increases with redshift. These results rely on a strongly parametrized model for the $\chi_\mathrm{eff}$ distribution and its correlation with $z$. Using a more flexible model for the correlation, \citeauthor{Heinzel:2023hlb}~\cite{Heinzel:2023hlb} showed that the correlation may be nonlinear, with width of the $\chi_\mathrm{eff}$ distribution increasing for $z<0.5$ but plateauing with large uncertainties for $z>0.5$---a possibility not allowed for with the parametric models of the previous analysis.

We test \textsc{PixelPop} on a simulated population from Ref.~\cite{Heinzel:2023hlb} that has such a correlation. In particular, the population of $\chi_\mathrm{eff}$ follows a truncated normal distribution on $[-1,1]$ for which the mean $\mu_{\chi_\mathrm{eff}}=0$ is constant but the standard deviation $\sigma_{\chi_\mathrm{eff}}$ increases in a nonlinear fashion with a cubic spline over increasing redshift. The spline has four nodes placed at $z=0,0.3,0.65,2.3$, over which the the standard deviation increases as $\ln\sigma_{\chi_\mathrm{eff}}=-3.5,-2,-1.5,-1.25$; see Tab.~\ref{tab:true_hyper}. We use \textsc{Pixelpop} to infer the two-dimensional merger rate density $\mathcal{R}(\chi_\mathrm{eff};z)$. The full dependence on the source parameters we consider is
\begin{align}
\mathcal{R} ( m_1 , q , \chi_\mathrm{eff} ; z )
=
\mathcal{R} ( \chi_\mathrm{eff} ; z )
p ( m_1 | \Lambda )
p ( q | m_1, \Lambda )
\, ,
\label{eq: model z chieff}
\end{align}
where we simultaneously infer the parameters of the primary mass $m_1$ and mass ratio $q$ distribution using the parametric \textsc{Power Law + Peak} model (jointly, this time, rather than just for $m_1$).

We show the inferred comoving merger rate density $\mathcal{R}(\chi_\mathrm{eff};z)$ in Figs.~\ref{fig:zchi_posterior_samples} and \ref{fig:zchi}, not evaluated at fixed $z$ because \textsc{PixelPop} infers the dependence on $z$ directly in this case. The former shows two specific draws from the population posterior---those with the maximum likelihood and maximum a posteriori values---as examples of how the inferred merger rate looks without being marginalized over posterior uncertainty. Compared to the prior draws in Fig.~\ref{fig: car prior samples}, there are overall structures driven by the information from the likelihood which mean these draws are less homogenous than the prior draws. The latter shows the posterior median and one-dimensional marginals with uncertainty. \textsc{PixelPop} again correctly finds the true correlation, namely that the effective spin distribution broadens nonlinearly as a function of increasing redshift. Since the true redshift distribution is relatively flat, we find that the overestimation of the merger rate near the edges of the parameter space is not as bad for the inferred marginal redshift distribution, but it is an issue for the narrow effective spin distributions at low redshifts. 

\begin{figure}
\centering
\includegraphics[width=0.9\columnwidth]{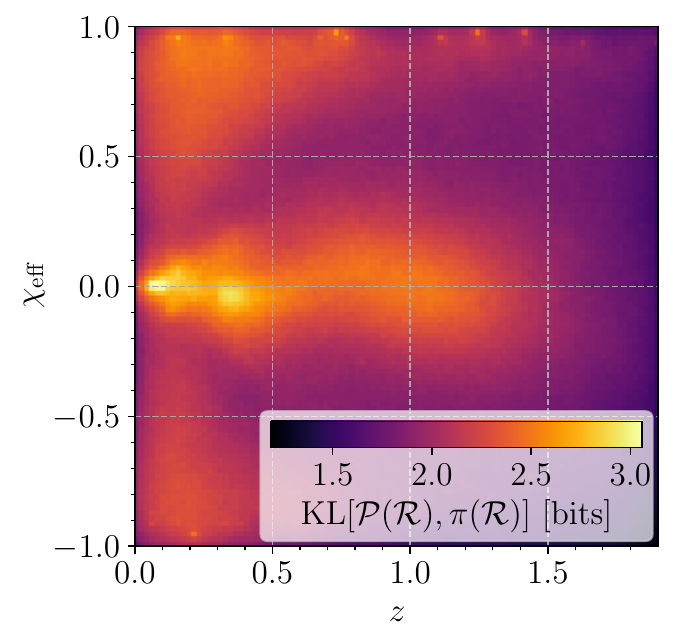}
\caption{
Information gained by the posterior $\mathcal{P}(\mathcal{R})$ inferred by \textsc{PixelPop} for each binned merger rate density over the effective prior $\pi(\mathcal{R})$ (see Sec.~\ref{sec: Quantifying information gain}) for the population with a nonlinear correlation between redshift $z$ and effective spin $\chi_\mathrm{eff}$. The information gain is quantified with the KL diverence $\mathrm{KL}[\mathcal{P}(\mathcal{R}),\pi(\mathcal{R})]$ in units of bits, with higher (lower) values shaded brighter (darker).}
\label{fig: kl z chieff}
\end{figure}

We compute the information gain in the merger rate posteriors over the effective prior using the KL divergence and plot the results in Fig.~\ref{fig: kl z chieff}. Similar to Fig.~\ref{fig: kl q chieff}, we see that the brightest pixels trace the detected sources, but that there are also larger KL divergences in regions where the true population density---and thus the number of detections---is low. This latter effect is far more pronounced in Fig.~\ref{fig: kl z chieff} than in Fig.~\ref{fig: kl q chieff} because the redshift of a source, which is a monotonic function of luminosity distance, much more strongly affects GW detectability than binary mass ratio or BH spin; the GW signal amplitude increases proportionally with decreasing luminosity distance. However, sources with positive $\chi_\mathrm{eff}$ are also more detectable on average than sources with negative $\chi_\mathrm{eff}$ due to the orbital hang-up effect \cite{Damour:2001tu, Campanelli:2006uy, Scheel:2014ina}: BH spins aligned (antialigned) with the orbital angular momentum act constructively (destructively) with the orbit to increase the total angular momentum and thus inspiral more slowly (quickly), therefore radiating more (less) energy through GWs. The combination of there being few such sources in the population and BH spins contributing to detectability at a subdominant level means this effect is not as visible in Fig.~\ref{fig: kl q chieff}. The effect in Fig.~\ref{fig: kl z chieff} is almost symmetric in $\chi_\mathrm{eff}$, meaning that being more detecbable is roughly just as informative as being less detectable. Similarly, we note that the KL divergence pixels in Fig.~\ref{fig: kl z chieff} can be just as bright in regions of no \textit{detections} but high \textit{detectability} as in regions of many detections, i.e., not detecting events can be just as informative as detecting events (though recall that the KL divergence is computed with respect to the effective ``informed prior'' defined in Eq.~(\ref{eq: effective prior single bin})). There are a few individual bright pixels in Fig.~\ref{fig: kl z chieff}, mostly along $\chi_\mathrm{eff}=0$, but these are just artefacts of the Monte Carlo approximations in Eqs.~(\ref{eq: single event likelihood estimator}) and (\ref{eq: Nexp estimator}).

\begin{figure}
\centering
\includegraphics[width=\linewidth]{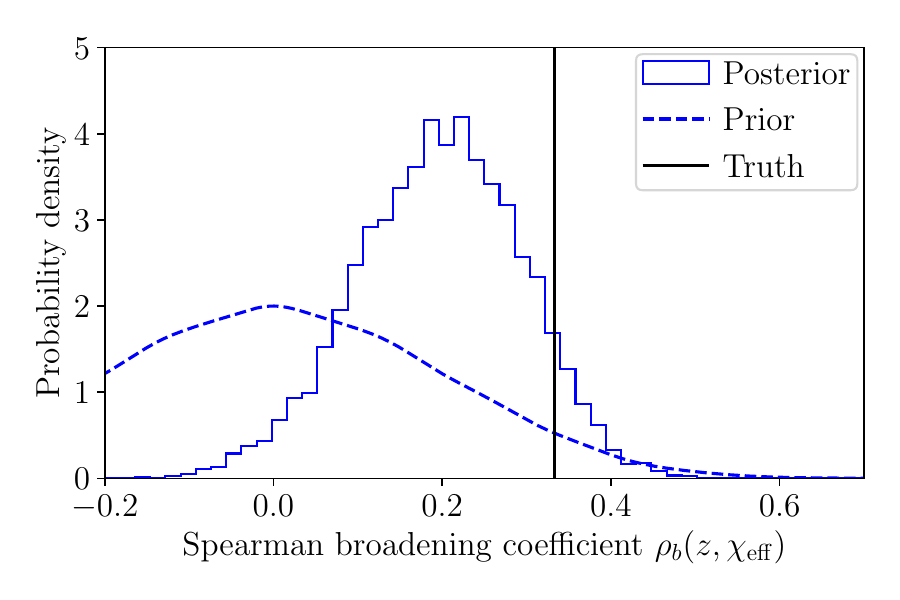}
\caption{Spearman broadening coefficient $\rho_\mathrm{b}(z,\chi_\mathrm{eff})$ from Eq.~(\ref{eq: broadening}) between the redshift $z$ and effective spin $\chi_\mathrm{eff}$, measured for $z\in[0,1]$, for the synthetic population in which the $\chi_\mathrm{eff}$ distribution broadens nonlinearly as a function of $z$. The solid line shows the posterior, while the dashed line shows the effective prior induced by the CAR model when marginalized over posterior of the CAR parameters; see Eq.~(\ref{eq: effective prior all bins}). The value $0.33$ for the true population is given by the vertical black line.}
\label{fig:Spearman_zchi}
\end{figure}

We now quantify evidence that the inferred distribution of effective spins $\chi_\mathrm{eff}$ broadens as a function of redshift $z$---which, indeed, the true population does. We use the broadening statistic we defined Eq.~(\ref{eq: broadening}) for $z$ and $\chi_\mathrm{eff}$, i.e., $\rho_\mathrm{b}(z,\chi_\mathrm{eff})$. However, as previously mentioned, the CAR model tends to naturally broaden in regions of uninformative data---such as large redshifts---so we would like to exclude this effect from our measurement of the broadening correlation. To do so, we restrict to $z\in[0,1]$. The resulting true value, effective prior from Eq.~(\ref{eq: effective prior all bins}), and posterior for $\rho_\mathrm{b}(z,\chi_\mathrm{eff})$ are plotted in Fig.~\ref{fig:Spearman_zchi}. We find $\rho_\mathrm{b}(z,\chi_\mathrm{eff}) = 0.20_{-0.17}^{+0.15}$ and that $\rho_\mathrm{b}(z,\chi_\mathrm{eff}) > 0$ with 97.4\% significance, consistent with the truth and with confidence in the existence of a broadening of the $\chi_\mathrm{eff}$ distribution of $z$. The posterior peak is somewhat offset from the true value (it lies at the $\approx93\%$ posterior quantile, or $\approx1.5\sigma$). This is generically expected for noisy observations but, following the discussion in Sec.~IIIB, measurements of the correlation and broadening coefficients are also influenced by the induced prior that disfavors values further away from zero. This is visible in Figs. \ref{fig:spearman_qchi} and \ref{fig:Spearman_zchi}, in which the tails of the posteriors (blue curves) are suppressed by the priors (dashed curves).

Altogether, our results suggest that with \textsc{PixelPop} we will be able to infer the presence of not just monotonic nonlinear correlations, but also nonmonotonic nonlinear correlations if they exist in the underlying population of binary BH mergers.

\subsection{No correlation}
\label{sec: simulated populations uncorrelated}

Finally, we consider a population in which the source parameters have no correlations; see Tab.~\ref{tab:true_hyper} for the true population parameters. Of the possible pairs of uncorrelated source parameters, we use \textsc{PixelPop} to infer the joint comoving merger rate density across redshift $z$ and effective spin $\chi_\mathrm{eff}$, as in Eq.~(\ref{eq: model z chieff}).

\begin{figure}
\centering
\includegraphics[width=1\linewidth]{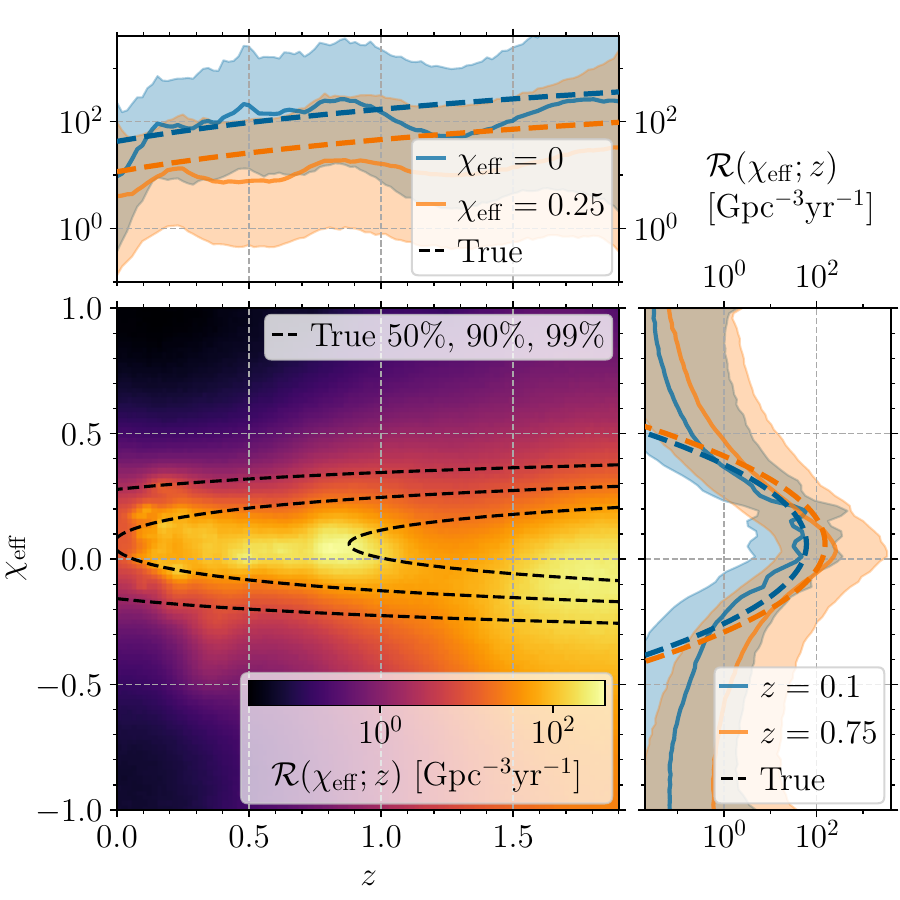}
\caption{Inferred comoving merger rate density $\mathcal{R}(\chi_\mathrm{eff};z)$ for a simulated population in which there is no correlation between any source parameters and \textsc{PixelPop} is used to measure the merger rate jointly over redshift $z$ and effective spin $\chi_\mathrm{eff}$. The central panel displays the two-dimensional posterior median, with lower to higher values shaded darker to brighter. The underlying true population is visualized with dashed black lines enclosing 50\%, 90\%, and 99\% of the distribution. The upper panel shows slices $\mathcal{R}(\chi_\mathrm{eff}=0;z)$ (blue) and $\mathcal{R}(\chi_\mathrm{eff}=0.25;z)$ (orange). The solid lines gives the posterior median and the shaded regions encloses the 90\% posterior credible regions. The dashed lines give the marginal distributions of the true population. Similarly, the right-hand panel shows slices $\mathcal{R}(\chi_\mathrm{eff};z=0.1)$ (blue) and $\mathcal{R}(\chi_\mathrm{eff};z=0.75)$ (orange).}
\label{fig:zchi_uncor}
\end{figure}

\begin{figure}
\centering
\includegraphics[width=0.9\columnwidth]{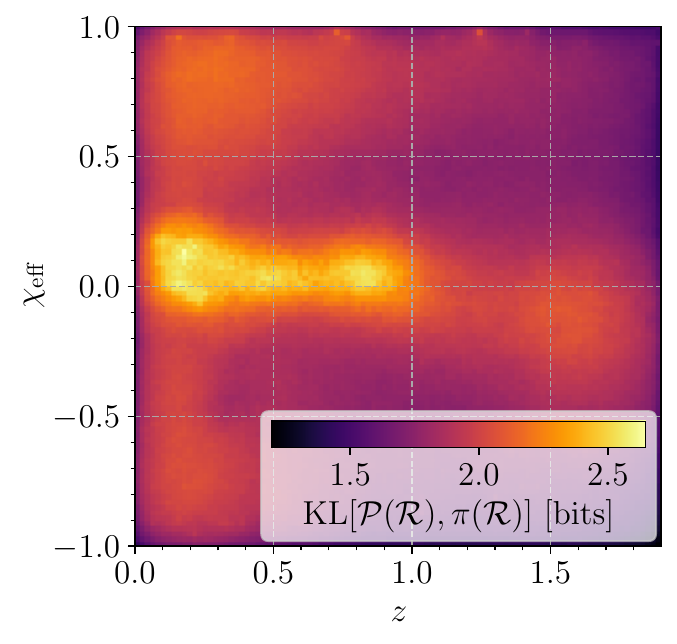}
\caption{Information gained by the posterior $\mathcal{P}(\mathcal{R})$ inferred by \textsc{PixelPop} for each binned merger rate density over the effective prior $\pi(\mathcal{R})$ (see Sec.~\ref{sec: Quantifying information gain}) for the population with no correlations between source parameters, and for which \text{PixelPop} is used to infer the merger rate joint as a function of redshift $z$ and effective spin $\chi_\mathrm{eff}$. The information gain is quantified with the KL diverence $\mathrm{KL}[\mathcal{P}(\mathcal{R}),\pi(\mathcal{R})]$ in units of bits, with higher (lower) values shaded brighter (darker).}
\label{fig: kl z uncor chieff}
\end{figure}

\begin{figure}
\centering
\includegraphics[width=\linewidth]{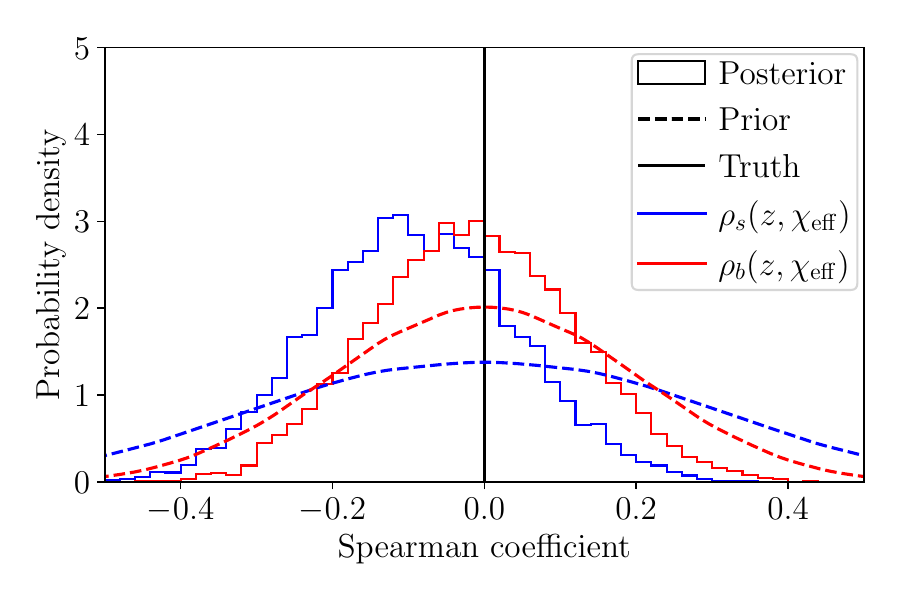}
\caption{Posterior (solid lines) and prior (dashed lines) distributions for correlation coefficients for the synthetic population in which the true distributions have no correlations. The true value of zero is marked with a vertical black line. The distributions in blue and red are for the Spearman rank correlation and broadening coefficients, $\rho_\mathrm{s}(z,\chi_\mathrm{eff})$ and $\rho_\mathrm{b}(z,\chi_\mathrm{eff})$, respectively, between redshift $z$ and effective spin $\chi_\mathrm{eff}$.}
\label{fig:Spearman_zchi_uncor}
\end{figure}

We present the inferred merger rate posterior in Fig.~\ref{fig:zchi_uncor}. The posterior has support for larger merger rates across all values of redshift regardless of the value of effective spin, as in the true population, and the inferred one-dimensional marginals are broadly consistent with the true distributions within the posterior uncertainty. The KL divergence between posterior and effective prior is given in Fig.~\ref{fig: kl z uncor chieff}. The information gain is qualitatively similar to Sec.~\ref{sec: Redshift and effective spin correlation}, with both detections and nondetections leading to informative regions across the parameter space. In Fig.~\ref{fig:Spearman_zchi_uncor} we show the inferred posterior for the Spearman correlation and broadening coefficients, $\rho_\mathrm{s}$ and $\rho_\mathrm{b}$. Both find no evidence for correlations, but they do not rule it out either as the posteriors are only midly constrained away from the priors toward favoring zero.

\subsection{Impact of unmodeled correlation}
\label{sec: unmodeled correlations}

\begin{figure*}
\centering
\includegraphics[width=0.49\linewidth]{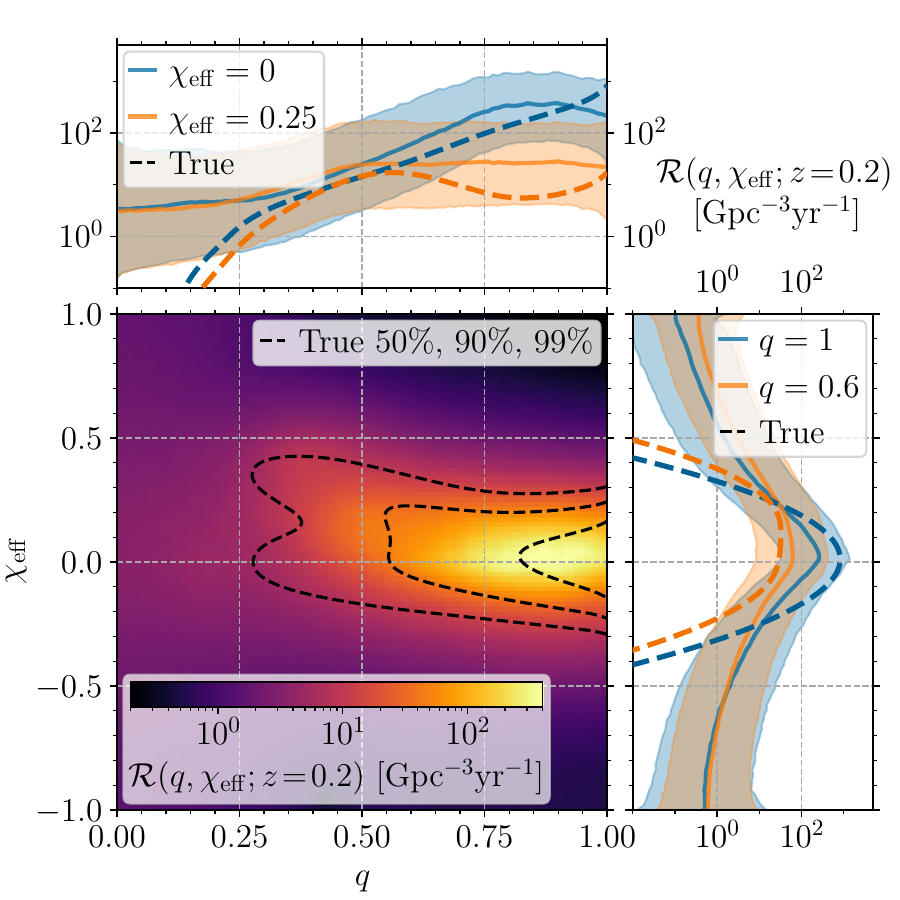}
\includegraphics[width=0.49\linewidth]{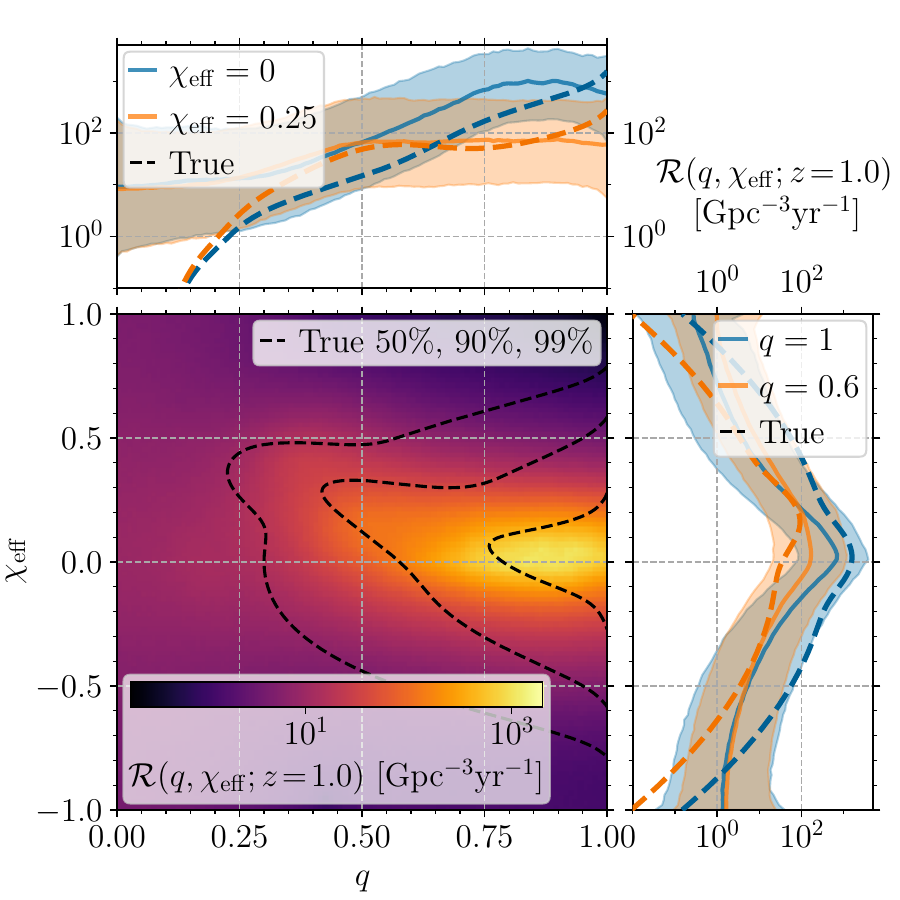}
\caption{\textit{Left.} Inferred comoving merger rate density $\mathcal{R}(\chi_\mathrm{eff}, q;z=0.2)$ at a fixed redshift $z=0.2$ for a simulated population in which there is a three-dimensional correlation between mass ratio, redshift and $\chi_\mathrm{eff}$. \textsc{PixelPop} is used to measure the merger rate jointly over mass ratio $q$ and effective spin $\chi_\mathrm{eff}$, but neglects the correlation with redshift. The central panel displays the two-dimensional posterior median, with lower to higher values shaded darker to brighter. The underlying true population is also evaluated at $z=0.2$ and visualized with dashed black lines enclosing 50\%, 90\%, and 99\% of the distribution. The upper panel shows slices $\mathcal{R}(\chi_\mathrm{eff}=0,q)$ (blue) and $\mathcal{R}(\chi_\mathrm{eff}=0.25,q)$ (orange). The solid lines gives the posterior median and the shaded regions encloses the 90\% posterior credible regions. The dashed lines give the marginal distributions of the true population. Similarly, the right-hand vertical panel shows slices $\mathcal{R}(\chi_\mathrm{eff},q=1)$ (blue) and $\mathcal{R}(\chi_\mathrm{eff},q=0.6)$ (orange).
\textit{Right.} The same as on the left hand side but the redshift evaluated at $z=1$.}
\label{fig: q chieff unmodeled correlation}
\end{figure*}

\begin{figure*}
\centering
\includegraphics[width=0.49\linewidth]{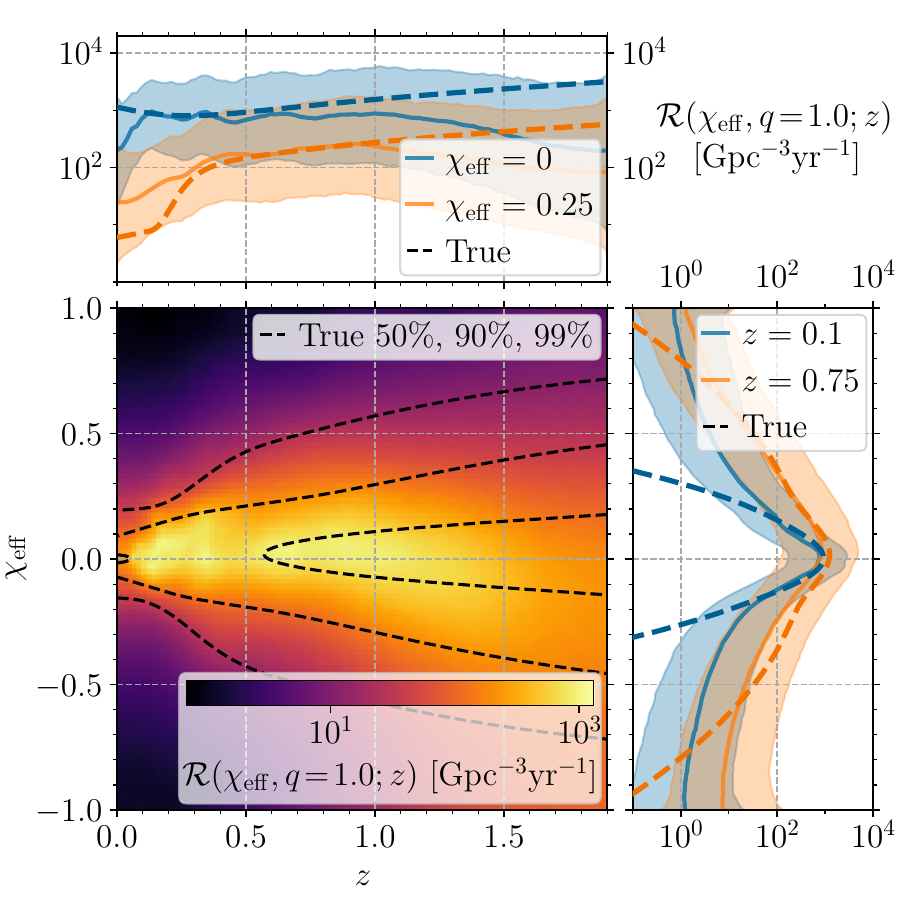}
\includegraphics[width=0.49\linewidth]{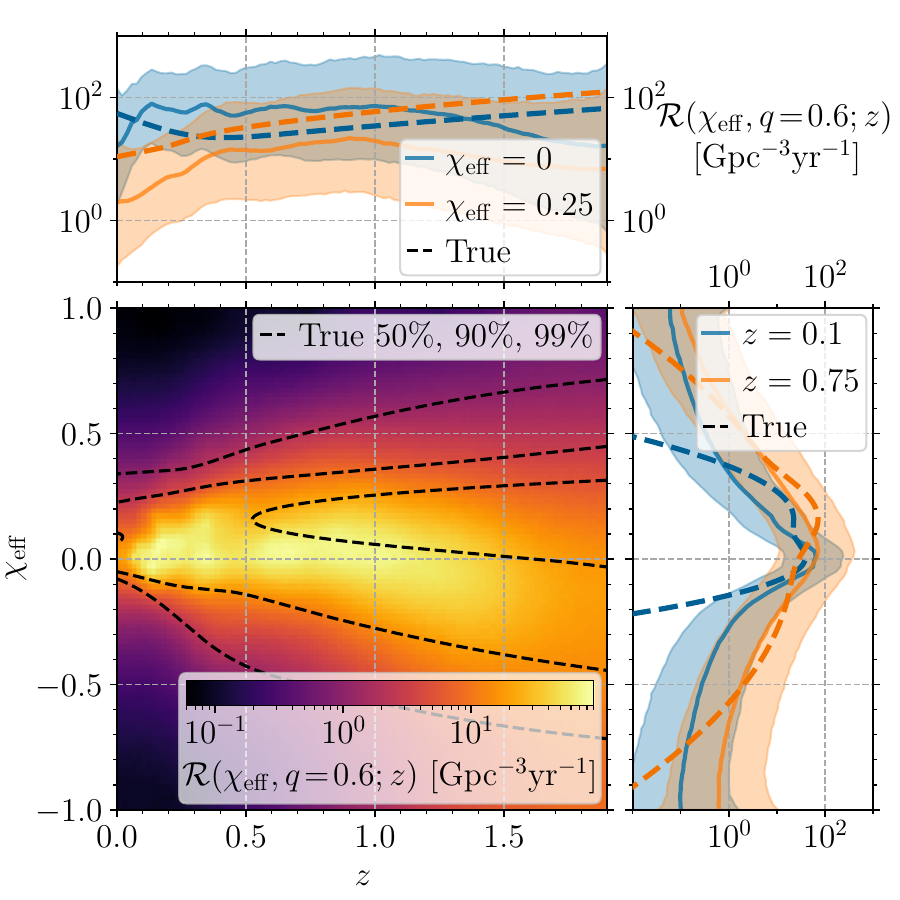}
\caption{\textit{Left.} Inferred comoving merger rate density $\mathcal{R}(\chi_\mathrm{eff},q=1;z)$ for a simulated population in which there is a three-dimensional correlation between mass ratio, redshift and $\chi_\mathrm{eff}$. \textsc{PixelPop} is used to measure the merger rate jointly over redshift $z$ and effective spin $\chi_\mathrm{eff}$, but neglects the correlation with mass ratio. The central panel displays the two-dimensional posterior median, with lower to higher values shaded darker to brighter. The underlying true population is also evaluated at $q=1$ and is visualized with dashed black lines enclosing 50\%, 90\%, and 99\% of the distribution. The upper panel shows slices $\mathcal{R}(\chi_\mathrm{eff}=0;z)$ (blue) and $\mathcal{R}(\chi_\mathrm{eff}=0.25;z)$ (orange). The solid lines gives the posterior median and the shaded regions encloses the 90\% posterior credible regions. The dashed lines give the marginal distributions of the true population. Similarly, the right-hand vertical panel shows slices $\mathcal{R}(\chi_\mathrm{eff};z=0.1)$ (blue) and $\mathcal{R}(\chi_\mathrm{eff};z=0.75)$ (orange).
\textit{Right.} The same as on the left hand side, where the mass ratio is fixed to $q=0.6$.}
\label{fig: z chieff unmodeled correlation}
\end{figure*}

We also consider a fourth population to study the impact of unmodeled correlations. This additional population contains an equal number of detected events from the two correlated populations above and hence represents a population with a trivariate correlation between $q$, $z$, and $\chi_{\rm eff}$. In the following analyses, we use \textsc{PixelPop} to model only a two-dimensional part of the full three-dimensional correlation. In our first analysis of this subsection, we model a $q$--$\chi_{\rm eff}$ correlation and assume redshift is independent (see Fig.~\ref{fig: q chieff unmodeled correlation}). Next, we model a $z$--$\chi_{\rm eff}$ correlation and assume $q$ is independent (see Fig.~\ref{fig: z chieff unmodeled correlation}).

Because there is a trivariate correlation, the model in both cases is incapable of capturing a correct representation of the true population. However, the flexibility afforded by \textsc{PixelPop} means that, within the statistical uncertainty, it can accurately recover the two-dimensional correlation between $q$ and $\chi_{\rm eff}$ at multiple values of redshift $z$, even as the nature of the correlation varies (c.f. the left and right hand panels of Fig.~\ref{fig: q chieff unmodeled correlation}). We note a similar result is found in the second analysis (Fig.~\ref{fig: z chieff unmodeled correlation}).

\begin{figure}
\centering
\includegraphics[width=\linewidth]{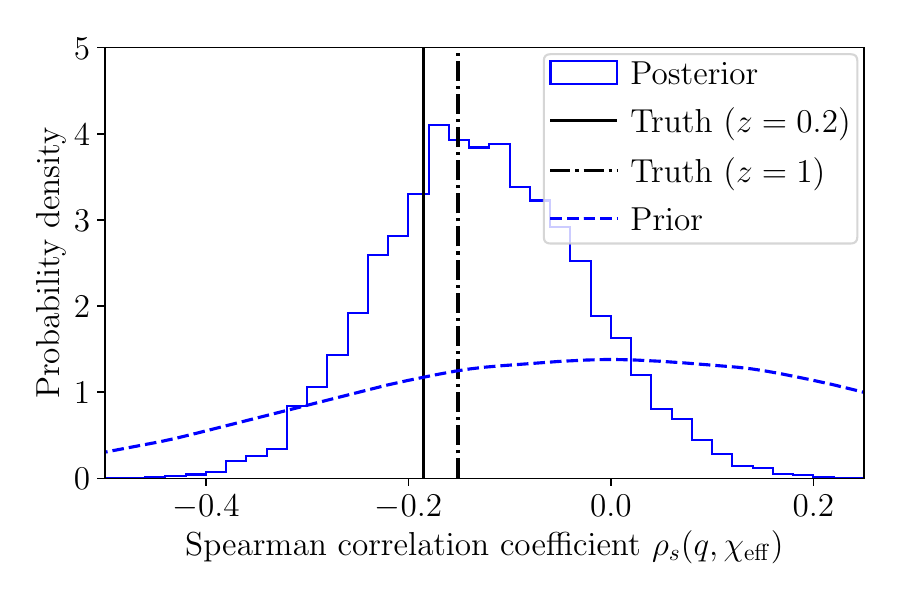}
\caption{Posterior (solid blue) and prior (dashed blue) distributions for the Spearman correlation coefficient for the synthetic population with a three-dimensional correlation, where \textsc{PixelPop} only models the two-dimensional mass ratio--effective spin distribution. The true value at redshift $z=0.2$ ($z=1.0$) is marked with a solid (dot-dashed) black line. The inferred Spearman correlation is invariant across different redshifts because we do not model a correlation with redshift.}
\label{fig: q chieff unmodeled spearman}
\end{figure}

For the $q-\chi_{\rm eff}$ correlated model, we show the inferred Spearman correlation coefficient $\rho_s(q,\chi_{\rm eff})$ in Fig.~\ref{fig: q chieff unmodeled spearman}. The inferred Spearman correlation coefficient is invariant at different redshift values as we do not model a correlation with redshift, while the true Spearman correlations do vary at different redshifts due to the underlying trivariate correlation. While we of course cannot recover the true trivariate correlation (the redshift distribution is forced to be independent), the \textsc{PixelPop} model uncertainties are large enough to account for the underlying correlation at a range of different redshifts.

\begin{figure}
\centering
\includegraphics[width=\linewidth]{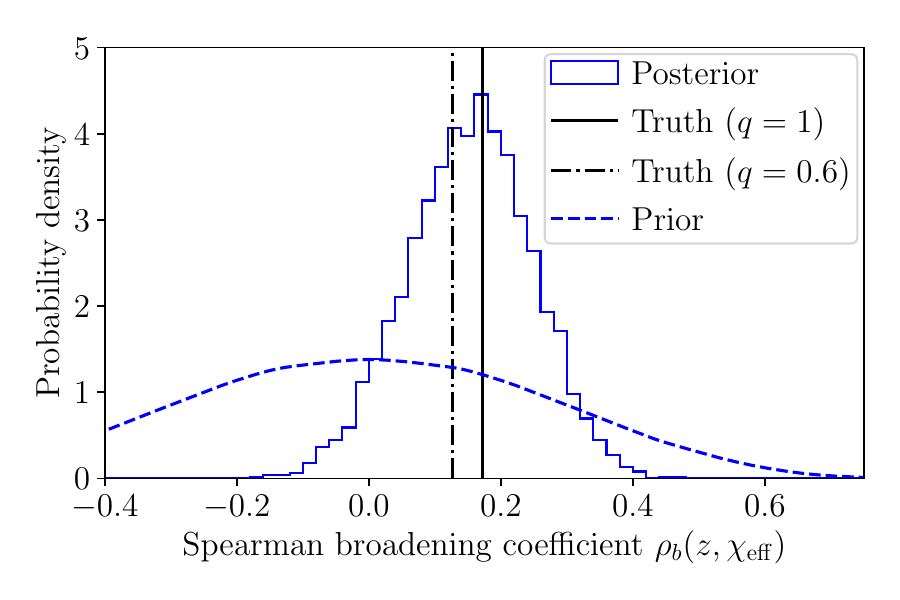}
\caption{Posterior (solid blue) and prior (dashed blue) distributions for the Spearman broadening coefficient for the synthetic population with a three-dimensional correlation, where \textsc{PixelPop} only models the two-dimensional redshift--effective spin distribution. The true value at mass ratio $q=1$ ($q=0.6$) is marked with a solid (dot-dashed) black line. The inferred Spearman broadening statistic is invariant across different mass ratios as we do not model correlations with mass ratio.}
\label{fig: z chieff unmodeled spearman}
\end{figure}

In Fig.~\ref{fig: z chieff unmodeled spearman}, we show the inferred Spearman broadening statistic $\rho_b(z,\chi_{\rm eff})$ compared at different values of mass ratio. Again, due to the large uncertainties inherent in our \textsc{PixelPop} approach, the inferred Spearman broadening statistic is consistent with the true values. We note that, by construction of the model which only allows for the bivariate correlation between $z$ and $\chi_{\rm eff}$, we cannot correctly identify the true three-dimensional correlation. 

\section{Conclusions}
\label{sec:conclusion}

In this paper, we presented \textsc{PixelPop}---a method for performing Bayesian nonparametric inference on the population of merging binary BHs detected with GWs. \textsc{PixelPop} directly infers the comoving merger rate density as a function of source parameter space with a binned representation, imposing a very weak smoothing prior that correlates each bins with other bins that are immediately adjacent. The CAR model at the core of \textsc{PixelPop} offers several advantages over similar methods---namely that it has more favorable computational scaling while simultaneously imposing weaker assumptions about the form of the merger rate. This allows us to perform inference on multidimensional GW populations with high resolution, for which we focus on bivariate parameter correlations in this work.

In a companion paper \cite{Heinzel:2024hva}, we apply \textsc{PixelPop} to the set of real LVK binary BH mergers and show that previously measured parameter correlations between BH spin, binary mass ratio, and redshift \cite{Callister:2021fpo, Biscoveanu:2022qac} cannot be confirmed without strong model assumptions until the GW catalog becomes more informative. In this work, we used \textsc{PixelPop} to test if we could confidently identify such correlations from mock GW catalogs drawn from simulated populations with complicated correlations between source parameters, using nonparametric measures of nonlinear and nonmonotonic correlations. Despite assuming almost nothing about the true population, \textsc{PixelPop} correctly inferred the merger rate densities and confidently identified parameter correlations with 400 GW detections in O4-like detector sensitivity. This implies that, while we cannot confidently conclude the real binary BH merger population is correlated between source parameters with the current catalog of $<100$ detections using flexible models like \textsc{PixelPop}, we will be able to do so in the near future---perhaps as soon as the end of O4. A downside to the model is that it becomes dominated by the CAR prior in regions of the parameter space in which there are few detections and the intrinsic GW detectability is low, meaning it overestimates the merger rate in those regions. This is an issue that is common to many flexible population models \cite{Golomb:2022bon, Edelman:2022ydv, Callister:2023tgi}, so we should be wary about such edges of the source parameter space.

Future work can use \textsc{PixelPop} to infer populations in even higher dimensions or just in a single dimension---the method described in Sec.~\ref{sec: Conditional autoregressive prior} applies to any number of dimensions, in principle.
Indeed, neglecting correlations in higher dimensions can lead to biased inference results when using population models that only allow for correlations between a subset of source parameters---an assumption made in all current GW population analyses that is likely invalidated by the astrophysical formation of GW sources. Extending our flexible model to higher dimensions will therefore be crucial for future work. In preliminary tests in Section~\ref{sec: unmodeled correlations}, we found that the large error bars on the merger rate inferred by \textsc{PixelPop} mean model misspecification such as this is more difficult to identify. 
However, because the models here only assume bivariate correlations, they are clearly unable to recover the trivariate correlations we simulated.

In three dimensions, 50 bins per dimension seems computationally feasible, i.e., $>10^5$ total parameters in the posterior of Eq.~(\ref{eq: posterior}). However, the HMC sampling time becomes slower and we are inevitably subject to the curse of dimensionality. It may be possible to gain computational efficiency via sparse matrix representations and limiting \textsc{PixelPop} to an intrinsic CAR model \cite{MORRIS2019100301}. An approach to overcome the large number of parameters may be to adaptively choose bin locations and resolution during inference based on regions of higher information gain, reducing the number of bins placed in regions of parameter space where the inferred merger rate posteriors are broad.

Other future applications can use \textsc{PixelPop} to flexibly model unknown populations, as \citeauthor{Cheng:2023ddt}~\cite{Cheng:2023ddt} showed that unmodeled subpopulations can bias the inferred contribution to the overall merger rate of other subpopulations. Furthermore, it is assumed in most population analyses that observations exceeding the detection threshold are real signals with absolute certainty, but noise transients can masquerade as GW signals. For instance, Refs.~\cite{Gaebel:2018poe, Roulet:2020wyq, Galaudage:2019jdx} attempt to include subthreshold GW triggers in population analyses, while Ref.~\cite{Heinzel:2023vkq} explicitly model the population of noise transients. We could use \textsc{PixelPop} as a more flexible model to encapsulate the rogue false-positive population of noise transients.

\textsc{PixelPop} can also be used to perform joint astrophysical and cosmological inference of GW populations. Ref.~\cite{Ezquiaga:2022zkx} showed that the so called ``spectral siren'' approach can yield accurate constraints on cosmological parameters; measurement uncertainties with current GW catalogs are modest, but will reach the percent level with future detectors \cite{Chen:2024gdn}. Refs.~\cite{Farah:2024xub, MaganaHernandez:2024uty} further demonstrated that strong parametric assumptions---which may lead to biased results---can be removed from the spectral siren method by using nonparametric Gaussian process priors in place of parametric population models, although current approaches assume there is no confounding astrophysical correlation. \textsc{PixelPop} can be used for agnostically modelling a potential astrophysical correlation as well as the cosmological parameters responsible for the redshifting mass spectrum. Outstanding issues that \textsc{PixelPop} could solve are: the computational cost of Gaussian process models with continuous covariance kernels, and; overconfident cosmological inference due to the presence of artificial features in the BH mass spectrum when using binned models with low resolution.

There remain some important technical problems. First, interpretability. For example, one may ask whether there are any binary BHs with negative effective spins $\chi_\mathrm{eff}$ \textit{at all} in the underlying population, as this implies constraints on possible formation scenarios. Previous analyses (e.g., Refs.~\cite{KAGRA:2021duu, Roulet:2021hcu}) have used targeted parametric models to understand exactly this and found that there is indeed evidence for negative $\chi_\mathrm{eff}$. However, such conclusions suffer from the use of strongly parametrized models and, therefore, model misspecification. From a nonparametric Bayesian perspective these conclusions are harder to make; our prior in Eq.~(\ref{eq: car prior}) \textit{requires} a nonzero merger rate across all parameter space, meaning one can never be fully convinced that there are exactly no mergers in certain parameter regions. However, we argue that this is a fair prior to use in the absence of confident theoretical models: how can we ever be convinced from observations alone that the population in a particular region is entirely devoid of sources? A nonparametric method like \textsc{PixelPop} can reliably place upper limits on the merger rate of binary BHs. There is not a one-size-fits-all approach for extracting astrophysical constraints from nonparametric results, but one possibility is hybrid models \cite{Edelman:2021zkw, Godfrey:2023oxb} that include some directly interpretable features on top of a more flexible underlying model or, conversely, flexible adjustments to an underlying parametric model.

A second major technical issue is inherent to the method for estimating the hierarchical likelihood in Eq.~(\ref{eq: total ln likelihood estimator}). The variance of the estimator, given in Eq.~(\ref{eq: likelihood variance}), scales poorly with the number of observations and the observation period \cite{Talbot:2023pex}, which requires us to impose an ad hoc regularization term; see Sec.~\ref{sec: Likelihood uncertainty}. This means we are a priori excluding regions of parameter space from the inferred posterior, due only to the likelihood estimation methods. New techniques that remove the Monte Carlo estimators in Sec.~\ref{sec: Likelihood estimation} may prove beneficial, e.g., Refs.~\cite{Talbot:2020oeu, Leyde:2023iof, Gerosa:2024isl, Mancarella:2024qle}.

Nonetheless, we have demonstrated the efficacy of \textsc{PixelPop} in inferring the underlying multidimensional merger rate of GW populations that have nontrivial correlations between source parameters---with high resolution, computational efficiency, and minimal model assumptions. As the number of GW observations increases, nonparametric methods will offer increasingly useful flexibility. A data release containing the results of our \textsc{PixelPop} analysis is available in Ref.~\cite{heinzel_2024_13176116}.

\section*{Acknowledgements}

We thank Amanda Farah, Jacob Golomb, Cailin Plunkett, Noah Wolfe, and the Rates and Populations LIGO working group for useful discussions and helpful comments.
We also thank the anonymous referee for raising interesting points and providing valuable feedback, which improved the quality of the manuscript.
J.H. is supported by the NSF Graduate Research Fellowship under Grant No. DGE1122374.
M.M. is supported by LIGO Laboratory through the National Science Foundation award PHY-1764464.
S.A.-L. is supported by the Thomas Frank fellowship fund at MIT.
J.H. and S.V. are partially supported by the NSF grant PHY-2045740.
This material is based upon work supported by NSF's LIGO Laboratory which is a major facility fully funded by the National Science Foundation.
The authors are grateful for computational resources provided by the LIGO Laboratory and supported by National Science Foundation Grants PHY-0757058 and PHY-0823459.

\bibliography{main}

\end{document}